\documentclass{aastex62}
\usepackage{amsmath}
\usepackage{bm}
\usepackage{relsize,exscale}

\published{2020 October 23}

\shorttitle{Fast Transit Computation Using Tabulated Stellar Intensities}
\shortauthors{Short et al.}

\begin{document}

\title{Fast Transit Computation Using Tabulated Stellar Intensities}

\correspondingauthor{Gur Windmiller}
\email{gwindmiller@sdsu.edu}

\author{Donald R Short}
\affil{Department of Astronomy, San Diego State University, 
5500 Campanile Drive,  San Diego CA 92182, USA}

\author{Jerome A Orosz}
\affil{Department of Astronomy, San Diego State University, 
5500 Campanile Drive,  San Diego CA 92182, USA}

\author{Gur Windmiller}
\affil{Department of Astronomy, San Diego State University, 
5500 Campanile Drive,  San Diego CA 92182, USA}

\author{William F Welsh}
\affil{Department of Astronomy, San Diego State University, 
5500 Campanile Drive,  San Diego CA 92182, USA}

\begin{abstract}
Limb darkening laws are convenient parameterizations of the stellar intensity
center-to-limb variation, and their use is ubiquitous in eclipse and transit
modeling. But they are not ``laws'' in any sense -- they are simple
approximations of the real intensity variations, and their limitations are
becoming more and more apparent as stellar atmosphere models improve and higher
precision data become available. When fitting eclipses and transit light
curves, one would ideally like to use model intensities that are based on
fundamental stellar parameters such as the mass, radius, and effective
temperature of the star, rather than a limb darkening law representation and
its coefficients. This is especially true when attempting to detect higher-order
effects such as planetary oblateness, rings, satellites, or atmospheres.
However, using model intensities requires numerically integrating many
small-area ``tiles'' on the model stellar surface(s) and this has
traditionally been too computationally expensive for general use.
Here we present a fast technique to compute light curves and the
Rossiter-McLaughlin effect that uses tabulated stellar models intensities.
This is a step in the development of tools that obviate the need for
limb darkening laws.
\end{abstract}


\section{Introduction} \label{sec:intro}

For centuries it has been known that the Sun's optical intensity decreases 
from the center of its disk to its limb.
This wavelength-dependent (and polarization-dependent) ``limb darkening'' 
effect\footnote{also known as the center-to-limb variation (CLV)}
is readily apparent in broad-band images of the Sun. 
The first quantiative measurement of limb darkening was made by Bouguer 
in 1748 \citep{Pettit1939}, and the many observational studies in the late 1800s 
and early 1900's were extremely valuable in the developement of 
our understanding of stellar atmospheres 
(e.g.\ \citealt{Schwarzschild1906}; \citealt{Milne1923}, \citealt{Milne1930}).
The ratio of the observed intensities relative to the center of the disk
was known as the ``law of darkening'', and can be expressed as the Laplace 
transform of the radiative source function for a plane parallel atmosphere 
(e.g.\ see \citealt{Przybylski1957}).
Thus the angular dependence of the intensity can be used to probe the source 
function with depth inside the star and constrain the temperature gradient
in the atmosphere.
While a boon to those interested in stellar atmospheres,
in the context of binary stars and exoplanets the limb darkening is in some 
sense a ``nuisance'' because it causes the eclipse profile to be 
deeper and rounded, instead of the trapezoidal shape that would result 
if the eclipse were a pure geometrical effect 
(e.g.\ see \citealt{Heller2019}).
Thus the limb darkening is correlated with the radius and inclination 
(impact parameter) of the orbiting bodies. So while usually of no intrinsic 
interest itself, getting the correct limb darkening is requisite to 
measuring the correct radius of the eclipsing star or transiting planet.
 \\

The intensity distribution across the stellar disk is non-trivial to
compute.  To overcome this computational bottleneck when modeling
eclipses, the standard practice has been to approximate the intensity
distribution with a limb darkening ``law''. The limb darkening law is
a function with only a few parameters (coefficients) that provides a
convenient way to estimate the intensities. The simplest law is the
linear law proposed by \citet{Schwarzschild1906}, but that is often too
crude of an approximation.  Since then, a variety of laws have been
proposed and used, for example the quadratic law (QL,
\citealt{Kopal1950}),
logarithmic law \citep{Klinglesmith1970}, square root law
\citep{Diaz1992}, and the Power-2 law \citep{Hestroffer1997}. These
are all two parameter laws; a 4-parameter ``non-linear'' law was
proposed by \citet{Claret2000} and can be reduced to the QL
when only two of the coefficients are used. In the case  of the Sun, where the
intensities can be directly measured, a 5th order polynomial is
sometimes used to match the intensity distribution
\citep{Neckel1994}. For other stars, the data are not of sufficient
resolution or quality to measure the limb darkening to such high
order, and the QL is very often used, especially for
exoplanet transit modeling. The 4-parameter non-linear law does
provide a better match to the stellar atmosphere model intensities,
but at the cost of two additional parameters. When using limb
darkening laws in practice, one can either adopt a limb darkening law
with coefficients that best match the atmospheric model intensities,
or use the data themselves to fit for and determine the
coefficients..   
Both methods have problems. \\

Adopting a limb darkening law and coefficients would seem at first to be the 
most sensible option, and this might be true if the laws actually matched the 
model intensities with high fidelity. But generally they do not, especially 
near the limb. A further complication arises in how the law should be matched 
to the intensities: to determine the coefficients of the law, should a 
least-square method or a flux conservation method be used 
(e.g.\ see Claret 2000 and references therein)? Also note that the 
best-fit coefficients are computed and tabulated for the center-to-limb 
intensities, but in practice, it is rare that the eclipsing body transits 
exactly through the center of the star. Thus the tabulated limb darkening 
coefficients do not correspond to the actual path the foreground body 
takes: the limb darkening law is not using the correct stellar intensities
(e.g.\ see \citet{Howarth2011}; \citet{Espinoza2016};
\citet{Morello2017};  \citet{Neilson2019}).
Finally, there remains the big question of whether the stellar model intensities 
are accurate. There are significant differences between plane-parallel and 
spherical models (for example \citealt{Hauschildt1999,Lester2008})
and more recent work has shown that the average intensities from 
the time-dependent 3D radiative hydrodynamic {\textsc{STAGGER}-grid} models 
(\citet{Magic2013},~\citet{Magic2015}) significantly differ from their simpler 
1D static counterparts and provide a better match to precision transit light curves
(\citet{Hayek2012}; \citet{Maxted2018}). \\

Empirically determining the limb darkening coefficients from the
eclipse data themselves has an apparent advantage: the added freedom
to adjust the model eclipse shape allows the model to better match the
observations.  However, this improved fit can be deceptive: the
derived radii can be significantly biased \citep{Morello2017}. This is
because, fundamentally, the limb darkening law approximations are not
identical to the stellar intensity drop off. The 4-parameter law
matches the intensities better than the other laws so the induced
radius bias is smaller, but it is still present (e.g.\ \citet{Morello2017} find 
absolute errors of $\sim$10-30 ppm in the measured transit depth, and 
Table 2 of \citet{Morello2020} lists a bias of a few ppm 
in relative transit depth depending on the wavelength range).
For comparison, these biases are a 
significant fraction of the observational uncertainties expected with the best telescopes, 
e.g., 30 ppm for {\it Kepler} (\citealt{Gilliland2011}), 
a few tens of ppm for {\it JWST} (\citealt{Beichman2014}), 
34 ppm for {\it PLATO} (\citealt{ESA2017}), 
and 10-100 ppm for {\it ARIEL} (\citealt{Tinetti2016}). 
It should be noted that when many transits are combined, 
such as can be done for the hot Jupiters, 
signals with an amplitude as small as a few ppm were measurable 
with ${\it Kepler}$ (e.g.\ \citealt{Esteves2015}).
Further complicating the issue is that the limb darkening coefficients are
correlated,{\footnote { The technique introduced by \citet{Kipping2013} 
allows much better sampling of the coefficient parameters when 
modeling data, but the correlation remains. Ad hoc methods such as 
solving for the sum and difference of the coefficients do help, 
but one would prefer a law in which the terms were intrinsically 
orthogonal.  }} leading to the degeneracy that many combinations
of coefficients can give nearly the same center to limb intensity
decrease over most of the disk. Thus a comparison of the
empirically-determined coefficients and the model-derived coefficients
is often misleading, although in principle it is possible to compare the 
intensity profiles instead of the coefficients.  \\

The conclusion one is led to from this discussion is that
parameterized limb darkening laws are convenient but inaccurate
approximations. At a time when the data quality were poor or when
plane-parallel stellar atmosphere models were the state-of-the-art,
the use of a limb darkening law was justifiable.  But with improved
stellar models
(e.g.\ ones that use better atomic input data and
spherical geometry rather than the
plane-parallel approximation), much higher quality observations 
 (e.g.\ ala {\it Kepler}), and investigations of more subtle phenomenon, one should
re-evaluate the use of limb darkening laws. \citet{Neilson2017} do
exactly this and conclude: ``We recommend the following changes to how
extrasolar planet transits are modeled: 1. Directly fit the CLIV
(center-to-limb intensity variation) from spherical model stellar
atmospheres to the observations.... shift from fitting limb-darkening
coefficients to fitting stellar properties such as effective
temperature, gravity, and stellar mass, which offers a way to
understand both the planet and its star to a new precision...''  \\

For decades there have been eclipse modeling codes that tile the stellar 
surfaces with ``pixels'' that are then directly summed to get the total
flux (e.g.\ the popular {\scshape{WD}} code of \citealt{Wilson1971}; 
{\scshape{LIGHT2}} by \citealt{Hill1979} and \citealt{Hill1993}).
Pixels that are covered by the foreground body during an eclipse are simply
not included in the summation.
But in these codes each pixel radiates like a blackbody (using the
appropriate local temperature after gravity darkening is included) 
and limb darkening laws are used to mimic the angular dependency.
There are a few codes that can use tabulated specific intensities from model 
atmosphere computations, for example 
{\scshape{BINSYN}} (\citealt{Linnell1994}; \citealt{Linnell2012}, \citealt{Linnell2013}), 
{\scshape{ELC}} (\citealt{Orosz2000}; \citealt{Wittenmyer2005}), 
{\scshape{PHOEBE2}} (\citealt{Prsa2005}; \citealt{Prsa2016}),
and most recently and noteably for exoplanets,
{\scshape{ExoTETHyS}} by \citet{Morello2020}.
However, the technique of integrating the intensities
is computationally expensive, especially for cases where there is
a large radius ratio that requires fine tiling of the star(s).
Hence these codes have generally not been used for fitting exoplanet 
transits, and instead the much faster limb darkening law approximations 
that have analytic solutions were employed instead (e.g.\ \citealt{Mandel2002}; 
and see the significant update by \citealt{Agol2019}).  \\

Here we present a new method that directly uses tabulated stellar
intensities, yet is fast, accurate, and general. It does not rely on
any special functional form for the limb darkening; indeed, it
completely eliminates the need for any limb darkening law
approximation. Using pre-computed tables of intensities corresponding
to stellar mass, temperature, surface gravity, and metalicity, the
method creates the eclipse profile for the given orientation of the
bodies. As such, the method enables a direct link between
state-of-the-art stellar atmosphere model intensities and
observational eclipse data. Of course, the accuracy of the treatment 
of the limb-darkening effect will now depend on the accuracy of 
the stellar models themselves. \\

There are two areas of exoplanet research where we expect this new
method will be potentially valuable: the study of planetary
oblateness and atmospheres. The first case arises because of the rapid
drop in intensity near the stellar limb. This intensity ``cliff'' is
poorly matched in all but the 4-parameter law. Somewhat surprisingly,
the effect of this cliff is most pronounced in the infrared, where the
limb darkening is often thought to be unimportant. In the infrared,
the reduced limb darkening means the intensity remains high from the
center of the stellar disk until close to the limb; thus the cliff is
larger and has a more dramatic effect. The ingress and egress phases
are most affected by the limb darkening law approximation, and this is
where the subtle effects of an oblate planet, a ringed planet, or
planetary satellites have the most effect. \citet{Neilson2017} show
that the error introduced by incorrect treatment of the intensity near
the limb can be as large as the amplitude of the signal produced by
planetary oblateness for the traditional quadratic limb-darkening law. 
For measuring exoplanet atmospheric composition,
transit spectroscopy relies on the wavelength-dependent change in the
measured planetary radius to discern the planet's atmosphere's
spectrum But there is a strong degeneracy between the transit depth
and the limb darkening, and wavelength-dependent errors introduced by
an inaccurate treatment of limb darkening can be as large as the
signal expected from a planet's atmosphere
(\citealt{Neilson2017}). Using stellar intensities directly instead of
any limb darkening law should help remove this degeneracy. 
In particular, we note that at the limb the line of sight passes nearly
tangentially though the star's atmosphere rather than radially, and thus
this ray is sampling gas that has only a small temperature gradient.
The source function is then nearly independent of optical depth.
Radiation from such an isothermal slab will, in general, have very weak
(or no) absorption lines.  But such behavior is not universal: in some cases, 
the lines will get limb brightened. \citet{Bitner2006} show a synthetic spectrum 
example of two Fe~I lines near 5522 \AA \ where one line decreases in 
strength toward the limb while the other increases.
Also, the core of a line can have different limb darkening than
the wings (e.g.\ see \citealt{Czesla2015}).
The crux of this is that limb darkening in spectral lines is very 
different than in the continuum, and transit spectroscopy is susceptible
to biases if this is not treated correctly (e.g.\ see \citealt{Yan2017}). \\

The method introduced here is based on previous work
(\citealt{Short2018}, hereafter ${\rm Paper~I}$) that employed Green's
theorem to compute eclipses of any number of overlapping spheres. We
briefly review that method in Section 2, then in Section 3 we develop
the new technique that replaces the handful of limb darkening law
approximations with tabulated intensities from stellar atmosphere
models.  
Section 4 describes the theoretical precision of the method, Section 5 
describes  the implementation of the method and how flux fractions through 
eclipse are computed, and Section 6 gives the numerical precision and speed of 
the method in practice.
In Section 7 we provide a set of examples using a mock 
hot Jupiter test case, and we summarize the results in Section 8. 
Four appendices are provided to explain in more detail the mathematical 
derivation and development of the technique. \\

\section{Method Overview}\label{sec:method_overview}

When computing the light curve of a partially eclipsed star, one needs
to evaluate a two-dimensional integral. The methodology of using
Green’s Theorem to reduce the integral over the surface of a sphere to
a line integral in a disc was pioneered by  \citet{Pal2012} and
extended in ${\rm Paper~I}$ to all limb darkening laws. In this work, we use the
theoretical results from ${\rm Paper~I}$ to replace the parametrized
analytic limb darkening laws by stellar atmosphere tables in the
calculation of light curves and the Rossiter-McLaughlin (R-M) effect. \\

As this work is based on the theoretical results of ${\rm Paper~I}$,
we will excerpt the salient portions of the theory from ${\rm
Paper~I}$ for the reader convenience. In the process of computing
observables for eclipsing binary star systems one encounters integrals
of the form
\begin{equation}\label{eq:gensurf}
\iint\displaylimits_{S_{\rm{vis}}} f(x,y,z)\cos(\gamma)dS 
\end{equation}
where $f=f(x,y,z)$ is a quantity such as intensity or rotational
radial velocity, $\gamma$ is the angle between the surface normal and
the direction to the observer (as was the case in ${\rm Paper~I}$),
$dS$ is the surface area element, and the integral is evaluated over
the portion of the surface visible to the observer. Throughout this
work, $(x,y,z)$ is a right hand coordinate system with $x$ and $y$
defining the plane of the sky (POS) and $z$ pointing towards the
observer. If we project opaque spheres onto the POS, we have
\begin{equation}\label{eq:gensurfproj}
\iint\displaylimits_{S_{\rm{vis}}} f(x,y,z)\cos(\gamma)dS = \iint\displaylimits_{\rm{D} _{\rm vis}} F(x,y) dA
\end{equation}
where the surface area element is related to the disk area element by
$dS = \sec(\gamma)dA$, and the integral is evaluated over the visible
part of the disk, $D_{\rm vis}$.  In the language of the exterior
calculus, the integrand $F(x,y)dA$ is a closed 2-form on the unit disk
(scaled by the disk radius).  By Poincar\'{e}'s Lemma, there exists a
1-form $P(x,y)dx+Q(x,y)dy$, the exterior derivative of which is $F$:
\begin{equation}\label{DPQdefine}
d\wedge\big[P,Q\big] \stackrel{\text{def}}{=} \frac{\partial Q}{\partial x} - \frac{\partial P}{\partial y} = F(x,y)
\end{equation}
where $d$ is the exterior derivative operator. Applying Green's theorem we obtain:
\begin{equation}\label{eq:greenapply}
\iint\displaylimits_{S_{\rm{vis}}}f(x,y,z)\cos(\gamma)dS = 
\iint\displaylimits_{D_{\rm{vis}}}F(x,y)dA =
\oint\displaylimits_{\partial D_{\rm{vis}}}\big[P,Q\big] \boldsymbol{\cdot} \big[x', y'\big] d\varphi
\end{equation}
where $x=x(\varphi)$ and $y=y(\varphi)$ is a right-hand oriented
parametric curve describing the boundary of the visible region, $x'$
and $y'$ are the derivatives of $x$ and $y$ with respect to $\varphi$,
and the integral is evaluated over the boundary of the visible disk
($\partial D_{\rm vis}$).  Thus, the evaluation of the visible surface
integral reduces to finding the description of the oriented parametric
curve and to expressing the 1-form $Pdx+Qdy$ in terms of simple
functions. \\

Equation 1-4 are general. For the purposes of computing a flux fraction 
through an eclipse one would require the detailed description of the 
parametric curves for any number of eclipsing bodies, and a specific functional 
form for $F(x,y)$, which in this case is the intensity projected on the POS. 
The detailed description of the parametric curve for any number of
eclipsing bodies is found in ${\rm Paper~I}$. We assume here the intensity 
on the POS is axisymetric and is a continuous function of the radius on the 
unit disk. Hereafter we will use the notation $I(r)$ to denote the intensity, which 
can be given by a parametrized limb darkening
law (see {\rm Paper~I}) or, as we show here, by a stellar-atmosphere
model.  For the R-M effect, the perturbation of the rotational radial
velocity field is given by
\begin{equation}\label{eq:deltaRV}
\delta RV = \frac{ \mathlarger{\iint}\displaylimits_{S_{\rm{vis}}}vI\cos(\gamma)dS }{\mathlarger{\iint}\displaylimits_{S_{\rm{vis}}}I\cos(\gamma)~dS}        
 \end{equation}
where $\delta RV$ is the RV perturbation, $v$ is the rotationally
induced radial velocity of the star.
Since the intensity $I(r)$ is axisymetric and is a continuous function on the unit disk,  
$I(r^2)$ is likewise a continuous function on the unit disk. Given this form, 
Appendix B in ${\rm Paper~I}$ specifies the expressions for the required 1-forms.
\begin{align}\label{eq:RMgfunc_summary}
\begin{aligned}
&d\wedge\bigg[-y\frac{G(r^2)}{r^2},x\frac{G(r^2)}{r^2}\bigg]=I(r^2) 
& &\text{(Flux fraction and R-M effect)} \\
&d\wedge\bigg[0,G(r^2)\bigg]=xI(r^2) & &\text{(R-M effect)} \\
&d\wedge\bigg[-G(r^2),0\bigg]=yI(r^2) & &\text{(R-M effect)} \\
\end{aligned}
\end{align} \\
where $G(r^2)$ is given by
\begin{align}\label{eq:RMgfunc_integral}
\begin{aligned}
G(r^2)=\frac{1}{2}\int\displaylimits_0^{r^2}I(z)dz \\ 
\end{aligned}
\end{align}
$G$ is an auxiliary function related to the ``flux'' enclosed within a radius $r^2$.
In this paper, we will assume that the function $F$ is given by a
table of intensity values that define a piecewise linear (PL)
function.  This table can originate from the values of a parameterized
analytic limb darkening law or from a
detailed model atmosphere computation.  Figure \ref{fig:method_overview}
provides a road map for the steps and analysis of this approach.  The
labeled arrows will be individually discussed.

\section{Development of the Method}\label{sec:method_development}

We begin with a stellar atmosphere model using spherically symmetric
geometry that for a given set of stellar parameters
(for example effective temperature $T_{\rm eff}$, ``surface'' gravity
$\log g$, mass, and metallicity) produces a table of center-to-limb
intensity variations integrated over well defined bandpasses.  Such tables were produced by
\citet{Neilson2013} for the standard Johnson-Cousins (BVRIHK) filters 
and the {\it Kepler} and CoRoT bandpasses. 
These tables sampled 1000 equally spaced points
of $\mu$, where $\mu=\cos(\theta)$ and $\theta$ is the angle formed
between a line-of-sight point on the stellar disk and the center of
the stellar disk. If we include the point at the star's center
$(\mu=1)$, which in the normalized case 
the intensity has a value of 1, we will have
a table of size 1001. The graph of such a table for the model stellar
atmosphere with parameters of effective temperature $T_{\rm eff}=5700
K$, gravity of $\log(g)=4.5$
a mass of $M=1.1 \,M_\odot$, and solar metallicity 
is shown in Figure \ref{fig:muVIntensity}.  
Note the cliff in the
intensities near the limb (e.g.\ $\mu\approx 0$).
This feature, which is a consequence of the spherical geometry and
thus is not present in the center-to-limb intensity variations
computed from model atmospheres that assume plane-parallel geometry, 
is not reproduced by most simple parameterized limb darkening
laws. \\

Our goal is to apply the theory of Appendix 
B in ${\rm Paper~I}$ (reproduced here for reader
convenience in Appendix \ref{sec:appendix_exderiv} for the flux 
computation and Appendix
\ref{sec:appendix_RMeffect} for the R-M effect) 
to these stellar atmosphere tables, expeditiously
producing light curves and models of the R-M effect. 
The first step is to change the
independent variable $\mu$ to $r^2$ using the relationship
$r^2=1-\mu^2$. This does not change the intensity values, but rather
re-indexes them by $r^2$. Since the independent variable in the new
table is decreasing, we will now reverse the order so that the new
independent variable is increasing.  By using linear interpolation
between the table values, we can construct a 
PL function, ${\rm Intensity} = I_{\rm PL}(r^2)$, which is a continuous
function (an example is shown in Figure \ref{fig:rsqVIntensity}). \\
To quickly produce interpolated values for a given $r^2$, we need to
identify the subinterval containing $r^2$ where $r_i^2 \le r^2 <
r_{i+1}^2$.  This index is given by
\begin{align}\label{eq:index_ceiling}
\begin{aligned}
i=N_\mu-{\rm ceiling}\big[(N_\mu-1)\sqrt{1-r^2}\big]\end{aligned}
\end{align}
where $N_\mu$ is the length of the table. The next step in 
the process of building the required 1-forms, is to evaluate 
the function $G(r^2)$ at the table values $r_i^2$, given by 
Equation (\ref{eq:RMgfunc_integral}). This is done recursively: 
\begin{align}\label{eq:G_recursive}
\begin{aligned}
G(r_{i+1}^2)=G(r_i^2)+\frac{1}{2}
\int\displaylimits_{r_i^2}^{r_{i+1}^2}I(r^2)d(r^2) 
\end{aligned}
\end{align}
Since the normalized intensity $I_{\rm PL}(r^2)$ is a PL function in
$r^2$, these integrals can be evaluated exactly by using the midpoint
rule (Gaussian 1-point). In addition, we can
compute the derivative of the integral $G(r^2)$ with respect to $r^2$ analytically:
\begin{align}\label{eq:PCHF_ddr}
\begin{aligned}
\frac{d}{d(r^2)}G(r^2)=\frac{1}{2}I(r^2)
\end{aligned}
\end{align}
Thus, we now have a new table with 3 rows. The first row consists of
the $r_i^2$ values, the second row contains the $G(r_i^2)$ values, and
the third row contains the derivative of the second row with respect
to the first row, or $\frac{1}{2}I(r_i^2)$. Note that the first row
together with twice the third row reproduces the original intensity as
a function of $r^2$. This implies that this step resulted in no loss
of information. The new table defines a smooth function $G_{\rm
  PCH}(r^2)$ using piecewise cubic Hermit (PCH) interpolation for
intermediate values (see Appendix
\ref{sec:appendix_tabledefined} for a thorough discussion). From Equation
(\ref{eq:RMgfunc_summary}), $G_{\rm PCH}(r^2)$ now defines the 1-forms
required by the R-M effect computation (Figure
\ref{fig:rsqVGrsq}). The final step is to compute $G(r^2)/r^2$,
which is needed to compute the
1-form (see  the top line in Equation
\ref{eq:RMgfunc_summary}) that is required for the light curve
computation. For values at each $r_i^2$, a simple division suffices,
except when $r_i^2=0$. The value at zero is defined by the limit as
$r^2$ goes to zero, and is computed in Appendix
\ref{sec:appendix_tabledefined}, giving a value of
$\frac{1}{2}I(0)$. The derivatives with respect to $r^2$ are computed
analytically, except at $r_i^2=0$: \\
\begin{align}\label{eq:fr2_ddr}
\begin{aligned}
\frac{d}{d(r^2)}\frac{G(r^2)}{r^2}= {1\over r^2}  \bigg[\frac{1}{2}I(r^2)-\frac{G(r^2)}{r^2}\bigg]
\end{aligned}
\end{align}
Here, the value is, again, defined by the limit as $r^2$ approaches
zero and is detailed in Appendix \ref{sec:appendix_tabledefined}. This
limit of the derivative has the value zero. \\

Since we know both the values and the derivatives of $G(r^2)/r^2$, 
this function 
may be expressed as a PCH function which we will denote as 
$G_{\rm PCH}(r^2)/r^2$ (see Figure
\ref{fig:rsqVGrsqDbyrsq}). From Equation (\ref{eq:RMgfunc_summary}),
$G_{\rm PCH}(r^2)/r^2$ now defines the last 1-forms required by both
the light curve and the R-M effect computation. This process converts
the given $I_{\rm PL}(r^2)$ into $G_{\rm PCH}(r^2)/r^2$ which defines
the 1-form required for the light curve computation. The top line of
Equation (\ref{eq:RMgfunc_summary}) is
\begin{align}\nonumber
\begin{aligned}
d\wedge\bigg[-y\frac{G(r^2)}{r^2},x\frac{G(r^2)}{r^2}\bigg]=I(r^2) \end{aligned}
\end{align}
In Appendix \ref{sec:appendix_tabledefined} we show that the values of
the exterior derivative at $r_i^2$ are equal to the values of the
intensity at those same $r_i^2$ values. Thus, there is no loss of
information as the original table can be reconstructed from $G_{\rm
  PCH}(r^2)/r^2$.  \\ 
In summary, 
to efficiently compute light curves from tables of model atmosphere
specific intensities,
those tables must be in the form of 
$G_{\rm  PCH}(r^2)/r^2$, which can be computed ahead of time.
Then, from Equation (\ref{eq:greenapply}) and Equation
(\ref{eq:RMgfunc_summary}), the path integrands are now a simple
product of a PCH function and sines and cosines. For a given value of
$r^2$, Equation (\ref{eq:index_ceiling}) provides the row index in the
table so searching the table is not required. \\

\subsection{Defining the Stellar Radius}

The definition of what the ``stellar radius'' is for a model
atmosphere with spherical symmetry merits some discussion.  What is
often done is to define the radial shell where $r=1.000$  to be where
the Rosseland mean optical depth $\tau_R$ reaches some critical value.
In the SAtlas models, used by \citet{Neilson2013} to produce
the grid of models we are using here, $\tau_R=2/3$ is set as the critical value
(\citealt{Lester2008}). Alternatively, the PHOENIX models use $\tau_{\rm std}=1$, 
where $\tau_{\rm std}$ is the optical depth in the continuum at $1.2 \mu m$ 
(\citealt{Hauschildt1999}).  The resulting model atmospheres are
insensitive to the precise choice of this critical value
(\citealt{Lester2008}).  As a practical matter, there is some light coming
from beyond the $r=1.000$ shell, so a small number of additional
shells are added.  For example, in the case of the PHOENIX models as
applied to Procyon, the outermost shell extended to $\approx 0.4\%$
beyond the $\tau_R=1$ shell (\citealt{Aufdenberg2005}).
The ``radiation field'' $I(\mu,\lambda)$ then refers to the angular
distribution of the intensities emerging from the outermost model
layer (\citealt{Aufdenberg2005}).  As a result of this
small extension, the $r=1.000$ radial coordinate defined to be where the optical
depth reaches the critical value will not occur at $\mu=0$ in a
spherically symmetric model atmosphere, but rather at the $\mu$ value
where the slope of the intensity profile is the steepest (see
\citealt{Espinoza2015} and cited references). Indeed, when fitting
limb darkening laws to intensity profiles from PHOENIX models,
\citet{Espinoza2015} put $\mu=0$ at the angle where the intensity
profile is the steepest and renormalized the angles.  By doing so,
the simple limb darkening laws can fit the renormalized profiles much
better. On the other hand, the details of what happens near the limb
are lost.  In our method, such a renormalization is not needed; 
$\mu=0$ is not reset to the location where the intensity drop off is steepest 
(nor is it where $\tau=2/3$ or $r=1.000$).  It corresponds to a radial coordinate that is 
very slightly greater than 1.000, in accord with the intensities tabulated 
by \citet{Neilson2013} from the SAtlas models. If one chooses $r^2$ to be 
the independent variable, as we do here, then
the integration effectively proceeds from the center of the apparent
disk (where $r^2=0$) outwards in the radial coordinate until the intensity drops to
zero, wherever that radial coordinate may be. Thus the use of $r^2$ as the radial coordinate effectively decouples the method from the precise definition of the stellar "radius".  \\

\section{Numerical Properties of the Table}\label{sec:numerprop_table}

While the Atmosphere Table Method (ATM) follows the theory described
in Appendices \ref{sec:appendix_exderiv} and
\ref{sec:appendix_tabledefined}, the implementation needs to be
checked. The first test was accomplished when it was shown that the
original stellar atmosphere table could be recovered without loss of
information from the table defining $G_{\rm PCH}(r^2)/r^2$ as
predicted by the theory. For the next
implementation test, we turn to the 
QL. As we have previously shown, the QL may be written in a dot
product form as detailed in ${\rm Paper~I}$. That form is given by:
\begin{align}
\begin{aligned} \label{eq:C_dot_Psi}
&I(\mu)/I_0 = \bm{C} \boldsymbol{\cdot} \bm{\Psi} \\
&\text{where } \\& 
 \bm{C}=\big[\big(1-(c_1+c_2)\big),(c_1+2c_2),-c_2\big] \text{~~and~~} \bm{\Psi}=\big[\bm{1},\mu,\mu^2\big]  \\
\end{aligned}
\end{align}
Here, $\bm{C}$ is a constant vector made from the standard QL
coefficients and $\bm{\Psi}$ is a set of basis functions defining the
law. The advantage of this form is that the function $G(r^2)/r^2$ has
a simple analytic form for each basis function, is independent of the
paths to be integrated over (system geometry), and is independent of
the law parameters. 
Equations (\ref{eq:greenapply}) and
(\ref{eq:RMgfunc_summary}) give the expression for the path
integrands, and these equations show the central role of the function
$G(r^2)/r^2$. Once the paths are given,  the path integrals can be
evaluated, again, independent of the law parameters. This allows for
modeling with several different wavelengths without re-computing the
paths or the path integrals. For most limb darkening laws we have a
listing, in  Table 1 of ${\rm Paper~I}$, of 
the analytic form of $G(r^2)/r^2$ for
the law's basis functions.  For the QL we
have:
\begin{align}
\begin{aligned} \label{eq:QLD_analytic}
&G(r^2)/r^2 {\rm ~for~} \bm{1} \text {~is given by~} \frac{1}{2} \\
&G(r^2)/r^2 {\rm ~for~} \mu \text{ is given by } (1-\mu^3)/(3r^2) \\
&G(r^2)/r^2 {\rm ~for~} \mu^2 \text{ is given by } (\frac{1}{2}-\frac{r^2}{4}) \\
\end{aligned}
\end{align}
The ATM also produces a $G_{\rm PCH}(r^2)/r^2$ for each of the QL
basis functions, which can be compared to the QL's analytic
forms. This comparison is independent of the paths (system geometry)
and the law parameters. The first QL basis function, $\bm{1}$,
results in the constant functions of the independent variable $r^2$
and thus $G_{\rm PCH}(r^2)/r^2$ is computed exactly to have the value
of $\frac{1}{2}$. The second QL basis function results in the graph,
shown in Figure \ref{fig:LinearBasisFrsq}, of the normalized intensity
with respect to $r^2$. Since the PL function, ${\rm Intensity}=I_{\rm
  PL}(r^2)$, shows an obvious curvature, the difference between this
PL approximation and the analytic form of the QL linear basis
function is $O(h^2)$, where $h$ is the subinterval size that depends
on the number of points in the table (see Figure
\ref{fig:LinearBasisGrsqDbyrsq}). The basis function $\mu^2$ results
in a linear function of the independent variable $r^2$ (Figure
\ref{fig:QuadBasisFrsq}), and hence $G(r^2)/r^2$ is computed exactly
to the precision of the machine. It is only the linear basis function
which produces any difference with the analytic form in the
development of the basis of 1-forms required for the construction of a
light curve using the ATM. If the table size is 1001, then the
difference between the model atmosphere's $G_{\rm PCH}(r^2)/r^2$ and
the QL's analytic form of $G(r^2)/r^2$ for the linear basis
function is smaller than $10^{-7}$. This, again, is independent of all
system geometries of spherical bodies and of all QL parameters. \\

\section{Computation of the Flux Fraction}\label{sec:implementation}

The next step in the analysis is to find the flux fraction.
This requires the computation of the path integral
given in Equation \ref{eq:greenapply}, the domain of which is the
boundary of the visible portion of the eclipsed body.
${\rm Paper~I}$ gives a thorough discussion on how to find the visible boundary,
and we summerize here a few key points. 
This boundary
consists of portions of the boundary of the obscuring disks and
portions of the boundary of the eclipsed disk. These bounding circular
arcs are parametrized by the subtended central angle. For the $i_{th}$
disk centered at $\big(X_{{\rm CTR}_i},Y_{{\rm CTR}_i}\big)$ we have:
\begin{align}
\begin{aligned} \label{eq:xrcos_yrsin}
&x=R_i \cos{\varphi_i}+X_{{\rm CTR}_i} \\
&y=R_i \sin{\varphi_i}+Y_{{\rm CTR}_i} 
\end{aligned}
\end{align}
where $R_i$ is the radius with center $\big(X_{{\rm CTR}_i},Y_{{\rm
    CTR}_i}\big)$ and the limits on $\varphi$ are
$\big[\varphi_0,\varphi_1\big]$. Switching to the coordinates of the
eclipsed body, $M$, in units of the eclipsed body's radius, we obtain:
\begin{align}
\begin{aligned}\label{eq:eclipsed_body_coord}
&x=\big[R_i \cos{\varphi_i}+\big(X_{{\rm CTR}_i}-X_{{\rm CTR}_M}\big)\big]/R_M \\
&y=\big[R_i \sin{\varphi_i}+\big(Y_{{\rm CTR}_i}-Y_{{\rm CTR}_M}\big)\big]/R_M \\
\end{aligned}
\end{align}
where we use the same parameter and parameter limits as
before. Finally, the orientation of these various parametrized arcs
must be consistent with the orientation of the boundary of the
eclipsed body. This requires that for those arcs derived from
occulting bodies, the limits of integration be switched, or that the
path integrals of those arcs be multiplied by $-1$. Thus,
\begin{align}\label{eq:limbdarkintegral_new1}
\begin{aligned}
\oint\displaylimits_{\partial D_{\rm{vis}}}\big[P,Q\big] \boldsymbol{\cdot} \big[x', y'\big] d\varphi  = \sum_{i=1}^{\rm{\#~of~Arcs}} (\pm1) 
\int\displaylimits_{\varphi_0}^{\varphi_1} \big[P,Q\big] \boldsymbol{\cdot} \big[x'_i, y'_i\big] d\varphi
\end{aligned}
\end{align}
where $\pm 1$ is the orientation. If we now focus on one of these
integrals using the top line of Equation \ref{eq:RMgfunc_summary} to
define $\big[P,Q\big]$, we obtain:
\begin{align}\label{eq:limbdarkintegral_new2}
\begin{aligned}
\int\displaylimits_{\varphi_0}^{\varphi_1}\bigg[-y\frac{G(r^2)}{r^2},x\frac{G(r^2)}{r^2}\bigg] \boldsymbol{\cdot} \big[x',y'\big]d\varphi = 
\int\displaylimits_{\varphi_0}^{\varphi_1}\frac{G(r^2)}{r^2}\big(-yx'+xy'\big)d\varphi
\end{aligned}
\end{align}
The first term on the right hand side 
is simply the PCH function described by the ATM and is
easily evaluated using Equation (\ref{eq:index_ceiling}) to find the
table row to use for computing the PCH interpolant for the given $r^2$
(See appendix \ref{sec:appendix_tabledefined}). The second term is
simply a function of $\sin(\varphi),\cos(\varphi)$ and a
constant. Therefore, the system's geometry will determine the various
path integrals to be evaluated. \\
 
We evaluate the definite integrals in Equation
(\ref{eq:limbdarkintegral_new2}) by Gaussian quadrature.  Following the
procedures outlined in ${\rm Paper~I}$, we divide each integration
interval into several ``panels'', with the number of panels given by
\begin{align} \label{eq:Nopt}
\begin{aligned}
N_{\rm opt}={\rm NINT}[(5p\Delta\phi+1)T_{\rm ATM}]
\end{aligned}
\end{align}
where $p$ is the ratio of the radii ($R_{\rm front}/R_{\rm back}$),
and $T_{\rm ATM}$ is a ``tolerance'' parameter. Note that this
definition of $N_{\rm opt}$ is similar, but not identical to the
definition from ${\rm Paper~I}$.  For clarity, we use the notation $T_{\rm
  ATM}$ for the tolerance parameter of the new method, and $T$ for the
tolerance parameter of the method in ${\rm Paper~I}$. After the number
of panels is set, $N_{\rm opt}$-point Gaussian quadrature is used up
to $N_{\rm opt}=64$.  If $N_{\rm opt}>64$, then we use $N_{\rm
  opt}/m$-point Gaussian quadrature on $m$ equal sub-intervals, where
$N_{\rm opt}/m$ is as large as possible without exceeding 64 (if
$N_{\rm opt}/m$ is not an integer, $N_{\rm opt}$ is set to the value
of the next multiple of $m$ that is larger than $N_{\rm opt}$). \\

\section{Numerical Speed and Accuracy of the new Method}\label{sec:speed}

Our ATM has two main sources of numerical uncertainty.
In the first case, the definite integrals in 
Equation (\ref{eq:limbdarkintegral_new2})
are
evaluated using Gaussian quadrature, so there is a
``quadrature
error'' associated with that.  In ${\rm Paper~I}$ we showed, 
using standard analysis techniques,
that the quadrature error gets smaller when the tolerance
(e.g.\ the number of integration panels) is increased, so one
can adjust the tolerance $T$ to achieve any target uncertainty
on the flux fraction   
(for example, using $T=128$ results in a (fractional)
quadrature error of $\approx 10^{-12}$). \\

In the second case, there is
a ``table error'' which has two contributions: (i) errors owing to
the fact that the
center-to-limb intensity values are tabulated at a
finite number of points, and  (ii) errors in the theoretical
model atmosphere intensities owing
to  inexact treatment of key physical processes and
uncertainties in the input atomic data and/or round-off errors
in the tabulated values.  \\

There are a few ways to estimate or compute the table error.  
A relatively trivial way is to compute tables of various lengths
using an analytic
limb darkening law.  Using these mock tables, we can compute
a light curve and compare that light curve to a light curve
computed using the method of ${\rm Paper~I}$ (using $T=128$ so that the
quadrature error is minimal).
Since the intensity profile is approximated by a PL function
$I_{\rm PL}(r^2)$,
the table
error should scale  $O(h^2)$ where $h=1/N_{\mu}$ is the subinterval
size between tabulated values of the angle $\mu$ (assuming
equally spaced values).  The scale factor in the error estimate
is related to the second derivative of
the function, namely $I^{\prime\prime}(\xi)$ for some
(unknown) $\xi$ in the interval.  Thus, the table error
will depend on the curvature of the center-to-limb intensity 
variations.  Since the common analytic limb darkening laws
(e.g.\ quadratic, square root, or logarithmic) have  similar
curvatures from center-to-limb, the magnitude of the table error
should not depend strongly on which analytic limb darkening law is
used.  Furthermore, since these analytic limb darkening laws produce 
center-to-limb variations that have less curvature than what is computed
from the model atmospheres that use spherical geometry, the measurements
of the table error for these analytic laws will yield a ``floor''
for the table error for a table of length $N_{\mu}$.
\\

When computing the flux fraction for the case of two overlapping
bodies, two important quantities are
$p$, which is the ratio of the radii
and $z$, which is
the separation of the centers on the POS, normalized to the
radius of the back body.  Following ${\rm Paper~I}$, we 
divided the ($p$,$z$) plane into a $2800\times 2800$
grid, where $0<p\le 3.5$ and $0<z\le 3.5$. 
For this exercise, we computed mock tables of various lengths using
the QL, 
the square root law, and the logarithmic law.
At each grid point in the ($p$,$z$) plane (excluding the
trivial cases of no overlap or total eclipses), we computed the flux
fraction using the method of ${\rm Paper~I}$ with $T=128$, and the new
method outlined here with
$T_{\rm ATM}=128$, and found the absolute value of the differences
$\Delta$.
For each table, we then recorded the median absolute difference
$\Delta_{\rm med}$ and the maximum absolute
difference $\Delta_{\rm max}$ in the grid.  Figure \ref{fig:plottableerror}
shows these differences for the various limb darkening laws
as a function of the table
size $N_{\mu}$.  The maximum differences are
well described by the power law function $\Delta_{\rm max}=0.08459N_{\mu}^{-1.9964}$
and the median differences are well described by the power law function
$\Delta_{\rm med}=0.01437N_{\mu}^{-2.0047}$.
Thus, as expected, the errors show the expected behavior
where a doubling of the table size reduces the error by a factor of
4.  In addition, the magnitude of the table error for a given
$N_{\mu}$ does not depend strongly on what analytic limb darkeing
law was used.  For a table length of $N_{\mu}=1000$ the table errors are
$\lesssim 10^{-7}$.  \\

Turning to the tables of model atmosphere intensities
computed by  \citet{Neilson2013}, we don't have an easy or
practical way to estimate errors in the intensities caused
by imperfect atomic data or inexact treatment of the
various physical processes.  If however, uncertainties
in the intensities were available, we can estimate the uncertainties
in the flux fractions. The details are
somewhat involved, and are given in Appendix 
\ref{sec:atm_err_analysis}.  
\citet{Neilson2013} published
918 tables over a range of temperatures, gravities,
and central masses. All of these tables have a length
of $N_{\mu}=1001$ when the intensity at the disk center 
($\mu=1$) is included.  If the uncertainties in the intensities
are normally distributed with a mean of zero and a standard
deviation of
$\sigma$, the maximum error in the flux fraction
for the {\em Kepler} band is about $0.077\sigma$ for the
\citet{Neilson2013} tables.  The intensities are
given to 7 digits, and we show
in Appendix \ref{sec:atm_err_analysis} that the 
round-off errors in the tables gives rise to
errors
on the flux fractions of
$\approx2\times 10^{-9}$.
Finally, in Appendix \ref{sec:atm_err_analysis}
we estimate the table errors in the flux fractions
owing to the finite lengths of the tables.
The maximum error estimate for the flux fraction
computed from
any table in the grid is $2.1\times
10^{-7}$
for the range of $0<p<3$ and $0<z<3$.  
This is about a factor of 2 larger than what one
would get using a mock table of length $N_{\mu}=1000$ computed from
an analytic limb darkening law.  \\

As was the case in ${\rm Paper~I}$, there is a tradeoff between the
speed of the algorithm and the accuracy of the light curves.  
Using the QL, we computed grids of
flux fraction differences using $T_{\rm ATM}=1$, 2, 4, and 8, and
table lengths of $N_{\mu}=1001$, 2001, 4001, and 8001. Figure
\ref{fig:error1} shows the error map for $T_{\rm ATM}=2$ and
$N_{\mu}=1001$.  The errors for total transits of small bodies
($p\ll1$ and $z<1-p$) are generally quite small ($<1\times
10^{-8}$). For all of the combinations of $T_{\rm ATM}$ and $N_{\mu}$,
we recorded the maximum difference and the median difference,
excluding the trivial cases of no overlap or a total eclipse.  These
results are tabulated in Table \ref{tab:moderror}, and Figures
\ref{fig:plothist1} and \ref{fig:plothist2} show a few frequency
distributions.  When $T_{\rm ATM}=1$, the maximum error and the median
error hardly change with increasing table length $N_{\mu}$. This
suggests that quadrature errors dominate the overall error. When the
tolerance is larger (for example $T_{\rm ATM}=4$), increasing the
table length results in smaller errors, but only up to a point. When
$T_{\rm ATM}\ge 2$ and $N_{\mu}\ge 1001$, the maximum error is
$<3.81\times 10^{-7}$ and the median error is $< 4.25\times 10^{-8}$. \\

We performed speed tests of the new algorithm using the same procedure
as we used \ in ${\rm Paper~I}$ (for these speed tests, the table was
precomputed). The results are shown in Figure \ref{fig:plottiming}.
The speed of the new algorithm with $T_{\rm ATM}=1$ is similar to the
speed of the algorithm of ${\rm Paper~I}$ with $T=2$. We used a table length
of $N_{\mu}=1001$, but since the algorithm does not require a search
of the table, the resulting speed is essentially independent of the
length of the table.

\section{Light Curve Comparisons}\label{sec:hotjup}

In order to compare transit light curves computed with 
the ATM outlined above with those computed using analytic
limb darkening laws, we chose the table from 
\citet{Neilson2013} with the stellar parameters of  $T_{\rm eff}=5700$,
$\log(g)=4.50$, and mass $M=1.1M_\odot$ as the starting point.
For this particular combination of mass and gravity the 
stellar radius is  $R=0.97663\,R_{\odot}$.  
We then placed a planet with a mass of $M_{\rm planet}=0.0011\,M_{\odot}$
and radius of $R_{\rm planet}=0.097633\,R_{\odot}$ in a circular
orbit with a period of 10 days.  For the light curves,
we computed transits in the {\em Kepler}, $B$, and $H$ bandpasses.
For each filter, a grid of models was computed using impact
parameters between 0 and 1 in steps of 0.01.  As discussed
above, the relative error of these light curves 
is $\lesssim 2\times 10^{-7}$
in units of the normalized flux. \\

The grids of light curves from the ATM were fit with models
that used the QL with the   
\citet{Kipping2013} reparameterization of the coefficients, the
Power-2 law using the \citet{Maxted2018} reparameterization of the
coefficients, and the Claret 4-parameter law, all computed using
the methods of ${\rm Paper~I}$.  {We did not add any noise 
to the grids of light curves. 
At each impact parameter, the limb darkening coefficients, 
the planet's radius, and the the inclination
were optimized using a downhill simplex
``amoeba'' algorithm.
Figures \ref{fig:FigAQuadKepDiff}, \ref{fig:FigBQuadBDiff}, and
\ref{fig:FigCQuadHDiff} show the light curve residuals 
for the QL models as a function 
the impact parameter 
for the {\em Kepler}, $B$, and $H$ bandpasses, respectively. 
In general, the residuals rise and fall several times across the
transit, and depend on the impact parameter of the input model. 
Figure \ref{fig:FigDQuad3Panel} shows, 
for each impact parameter in the grid,  
the optimal QL coefficients, 
the relative error in the planet radius, 
and the relative error in the inclination.
The optimal limb darkening coefficients depend on the impact
parameter---this has been previously discussed in 
\citet{Neilson2017}, \citet{Howarth2011}, \citet{Kipping2011a}, and 
\citet{Kipping2011b}. 
Likewise, the relative error 
in the planet's radius and the relative error in the inclination  
also depend on the
impact parameter.  Furthermore, these
relative errors also vary with bandpass.  
In the case of the $H$ bandpass,
the maximum flux difference from Figure \ref{fig:FigCQuadHDiff} is about 
20 ppm (1 part in 50,000).
More importantly,
the relative error in the planet's radius for impact parameters 
$< 0.85$ is about 0.2\% for the worst case
as shown in Figure \ref{fig:FigDQuad3Panel}.
To better understand the behavior of this radius bias,
we used a nested sampling algorithm 
\citep{Skilling2006} to estimate the uncertainties 
in the fitted parameters.  
This was done for the models at impact parameters 0.0 through 0.8 
in steps of 0.1, and also for an impact parameter of 0.85.  
The nested sampling solutions are shown as the horizontal 
set of points
in Figure \ref{fig:FigDQuad3Panel}.
Notice how the uncertainties in the 
radius
grow larger as the impact parameter increases.\\

Figures \ref{fig:FigEPowKepDiff}, \ref{fig:FigFPowBDiff}, and
\ref{fig:FigGPowHDiff} show the results for the Power-2 law.
Note that the range of the color scale
has been reduced by a factor of 4 from the QL cases, and
that the $B$ and $H$ band graphs exhibit some noise. This noise would
imply the Power-2 light curve models are within a digit of the
light curve models generated by the ATM. The $B$ and $H$ band model
differences with the corresponding ATM models are within 
ten
parts per million. 
Similar to the case of the QL, 
Figure \ref{fig:Pow2LawParamsAndErrors} 
shows how the Power-2 law coefficients, relative radius error, and 
relative inclination error vary with impact parameter.  
The worst case for 
the relative planet radius error is $\sim$ 0.1\%. 
The Power-2 law provides an improvement of roughly a factor of two 
in the radius error over the QL.
For brevity, we omit the analogous two-dimensional images showing 
the transit residuals for the Claret 4-parameter law
and just present the fitting results.
Figure \ref{fig:Claret3PanelNov28} shows
the four limb darkening coefficients for the {\it Kepler} band, 
the error in the planet radius, and the error in the inclination.
As seen in the QL and Power-2 law, these are all dependent on the
impact parameter.  
Despite only small deviations in the transit light curve (few ppm), 
the error in the derived planet radius remains much larger, on 
the order of $\sim$0.1\%, due to various trade-offs
among the correlated parameters when optimizing to get the best fit.
For the reader's convenience we provide a {\sc Matlab} 
demonstration code that produces a synthetic light curve of a transiting 
mock planet by using the ATM and the method outlined here and in 
${\rm Paper~I}$. The code can be downloaded 
from https://doi.org/10.5281/zenodo.3473851
\citep{Short2019b}.

\section{Summary}\label{sec:summary}

While limb darkening laws are a convenient parameterization of the limb darkening 
phenomenon, they can sometimes be inaccurate representations of the stellar 
intensities.
This is particularly true near the limb of the star, where the intensity rapidly 
drops to zero. In the era of ultra-precise photometry from the {\it Kepler} 
Mission, the use of limb darkening law approximations to match the light curves 
can lead to systematic biases in the derived model parameters, such as planet 
radius, inclination (impact parameter), and many other correlated parameters. 
As shown in previous studies, it would be advantageous to use an actual model 
stellar atmosphere to give the specific intensities instead of relying on limb 
darkening law approximations. In general, this has not been done because of the 
severe computational burden. In this work, we provide a fast method to allow the 
use of tabulated model stellar atmosphere intensities, thus removing the need for limb 
darkening laws (and solving for their coefficients).

Our methodology is an extension of the work presented by ${\rm Paper~I}$ 
\citep{Short2018}, where we apply Green's Theorem to solve the two dimensional integral 
needed to compute the observed flux during an eclipse or transit.
Equations \ref{eq:RMgfunc_summary} and \ref{eq:RMgfunc_integral} 
give the one-forms needed to compute the intensity and R-M effect, and
in the present work we show how the function $G(r^2)$ in these equations 
can be found numerically given a table of specific intensities from a
model atmosphere calculation. More specifically, 
the input table of radii and intensities is recast as a piecewise linear 
function, and from that tables of $G(r^2)$ and $G(r^2)/r^2$
are constructed. These tables uniquely define a 
piecewise cubic Hermite function which can then be used for 
interpolation for any intermediate values of $r$.
The conversion from the model atmosphere table (\ref{eq:table_of_values})
to the form of the tables used in computing the values of the integrand
(\ref{eq:new_new_table_of_values}) is done once. It is reversible, implying 
no loss of information. Any error in the computed light curve stems 
mainly from the sizes of the steps (gaps or intervals) in the tabulated 
intensities (and of course from the limitations inherent in the stellar 
atmosphere models themselves).
For a two-body eclipse, the method is only a factor of $\sim$ 2 slower than 
the \citet{Mandel2002} code, but the inclusion of actual stellar physics and the 
{\it a priori} known level of precision more than compensate for the additional compute time.
The light curve or R-M radial velocity curve for any number of overlapping 
spherical bodies can be computed. \\

Using the tabulated specific intensities computed with the ATLAS stellar 
atmosphere models from \citet{Neilson2013}, we make a direct comparison 
of our atmosphere table method with 
the traditional quad law and Power-2 limb darkening laws for a hot Jupiter 
transit. Noting that the best-match quad law coefficients depend on the impact 
parameter, differences of several tens of ppm are present in the transit.
The derived planet radius can be systematically biased by as much as 
$\sim$  500 ppm in the H-band, depending on the impact parameter. 
While this is a small bias, such errors can be problematic as photometric 
precision continues to improve and as demands on the data increase, e.g., 
when considering higher-order effects such as planet oblateness, rings, 
satellite transits, etc. An area where our method may be particularly valuable 
is transit spectroscopy. A transit depth is wavelength-dependent due to both 
the planet's atmosphere and the stellar limb darkening.  
By eliminating the use of parameterized limb darkening law approximations, we (i) take advantage of the full knowledge of the star's intensity distribution (which can strongly vary with wavelength) from the stellar atmosphere models, and (ii) remove the degeneracy between planet radius and the empirically-constrained limb darkening coefficients because there are no coefficients in the model - the limb darkening is entirely set by the stellar mass, radius, temperature, and metallicity.
Because the limb darkening has no free parameters, the accuracy of 
the treatment of the limb darkening now depends entirely on the accuracy of 
the stellar models used. \\

This current paper describes how to use tabulated stellar atmosphere model 
intensities instead of limb darkening laws, under the assumption that such a 
table exists for the desired stellar parameters. 
Future work involves being able to precisely and efficiently interpolate the 
input stellar atmosphere tables across values of $T_{\rm eff}$, $\log(g)$ and 
metallicity.

\appendix

\section{Anti-Exterior Derivatives For Radially Defined Function}\label{sec:appendix_exderiv}

Assume $f=f(r^2)$ is a continuous function on the unit disk (every limb darkening is of 
that form). Since the 2-form $f(x,y)dxdy$ is a closed form on the unit disk, Poincar\'{e}'s Lemma
asserts that there exists a 1-form $P(x,y)dx+Q(x,y)dy$ such that
\begin{equation}\label{eq:DPQpartial}
d\wedge\big[P,Q\big]=\frac{\partial Q}{\partial x}-\frac{\partial P}{\partial y}=f(x,y)
\end{equation}
Try $P(x,y)=-yF(r^2),~~Q(x,y)=xF(r^2)$ for some smooth function $F$ 
on the unit disk. Then
\begin{align}\label{eq:DPQpartial_or}
\begin{aligned}
&\frac{\partial Q}{\partial x}-\frac{\partial P}{\partial y}=F(r^2)+2x^2\frac{dF(r^2)}{dr^2}+F(r^2)+2y^2\frac{dF(r^2)}{dr^2} \\
&\rm{or}\\
&2F(r^2)+2r^2\frac{dF(r^2)}{dr^2}=f(r^2) \\
&{\rm or}\\
&\frac{2d\big(r^2F(r^2)\big)}{dr^2}=f(r^2)
\end{aligned}
\end{align}
integrating $f=f(z)$ with respect to $z$ where $z=r^2$, from $0$ to $r^2$, gives
\begin{align}\nonumber
\begin{aligned}
2r^2F(r^2)=\int\displaylimits_0^{r^2}f(z)dz
\end{aligned}
\end{align}
or
\begin{align}\label{eq:integrateFr2}
\begin{aligned}
F(r^2)=\frac{\mathlarger{\int}\displaylimits_0^{r^2}f(z)dz}{2r^2}
\end{aligned}
\end{align}
Thus the 1-forms are $P=-yF(r^2),~~Q=xF(r^2)$, the exterior 
derivative of which is $f(r^2)$. \\

In the case of the R-M effect, all of the terms in the numerator of
Equation (\ref{eq:deltaRV2}), using any of the limb darkening laws,
have the 2-form $g=xf(r^2)$ or $h=yf(r^2)$ on the unit disk.
Try
\begin{align}\label{eq:P1P2Q1Q2}
\begin{aligned}
&P_1(x,y)=0 &Q_1(x,y)&=G(r^2) \\
&P_2(x,y)=G(r^2) &Q_2(x,y)&=0 \\
\end{aligned}
\end{align}
Therefore,
\begin{align} 
\begin{aligned} \nonumber
&\frac{\partial P_1}{\partial y}=0 \\
&\frac{\partial Q_1}{\partial x}=\frac{dG(r^2)}{dr^2}(2x)
=2x\frac{dG(r^2)}{dr^2}=g(x,y)=xf(r^2) \\
&{\rm or}~ \frac{dG(r^2)}{dr^2}=\frac{1}{2}f(r^2) \\
\end{aligned}
\end{align}
which we then integrate:
\begin{align} \nonumber
\begin{aligned}
&G(r^2)=\frac{1}{2}\int\displaylimits_0^{r^2}f(z)dz \\
\end{aligned}
\end{align}
Hence
\begin{align} \nonumber 
\begin{aligned}
&d\wedge\big[0,G(r^2)\big]=xf(r^2) ~\rm{and}\\
&d\wedge\big[-G(r^2),0\big]=yf(r^2)\\
\end{aligned}
\end{align}
In summary,
\begin{align}\label{eq:RMgfunc_integral_appendix}
\begin{aligned}
G(r^2)=\frac{1}{2}\int\displaylimits_0^{r^2}f(z)dz \\ 
\end{aligned}
\end{align}
where $f=f(r^2)$ is a continuous function on the unit disk. Then,
\begin{align}\label{eq:RMgfunc_summary_appendix}
\begin{aligned}
&d\wedge\bigg[-y\frac{G(r^2)}{r^2},x\frac{G(r^2)}{r^2}\bigg]=f(r^2) \\
&d\wedge\bigg[0,G(r^2)\bigg]=xf(r^2) \\
&d\wedge\bigg[-G(r^2),0\bigg]=yf(r^2) \\
\end{aligned}
\end{align} \\
We note, again, that Equations (\ref{eq:RMgfunc_summary_appendix}) 
explicitly give the 1-forms $[P, Q]$ for the flux calculation for 
any limb darkening given by
\begin{align} \nonumber
\begin{aligned}
I(\mu)/I_0=f(r^2)=f(1-\mu^2)
\end{aligned}
\end{align}
(where $f$ is a continuous function on the disk $D$), and for the
1-forms $[P, Q]$ for the R-M effect based on that limb darkening. The
specific form of the integral $G$ (Equation
\ref{eq:RMgfunc_integral_appendix}) will determine if $[P,Q]$ can be
expressed in closed form, by a special function, or will require
numerical evaluation using Gaussian quadrature. \\

\section{Anti-exterior Derivatives for the 
Rossiter-McLaughlin Effect}\label{sec:appendix_RMeffect}

Following the development of the R-M effect from Section 4.2 of ${\rm
  Paper~I}$, define the rotation axis of the star in the dynamic
coordinate system $(x,y,z)$, with the observer on the positive
$z$-axis, as defined by Equation \ref{eq:gensurf}. Note that this
differs from the system first defined \citet{Hosokawa1953} and then
used again by \citet{Gimenez2006a}. Since their emphasis was on simple
binary systems and planetary transits, their $y$-axis was simply the
projection of the orbital pole on the POS. In this work we do not
assume a simple two-body system but rather a multi-body system in
which
\begin{align}
\begin{aligned}\label{eq:ThetaPhi_definition}
& \Theta_{\rm rot} \equiv \text{~~angle in the $(x,y)$ 
plane from the $y$-axis} \\
& \Phi_{\rm rot} \equiv \text{~~ angle from the $z$-axis colatitude}
\end{aligned}
\end{align}
If the axis of rotation of the star is the $z$-axis, then the right-hand 
surface velocity field 
on the unit sphere is given by 
\begin{align}
\begin{aligned}\label{eq:angle_unitsphere}
v(x,y,z)=\omega (-y,x,0)
\end{aligned}
\end{align}
where $\omega$ is the angular velocity in radians per day. Note that
$\lVert (-y,x,0)\rVert$ is the distance to the axis of rotation (the
$z$-axis). Using rotation transformations (orthonormal matrices with
determinants equal to 1), move the $z$-axis to the rotation axis
described by $(\Theta_{\rm rot},\Phi_{\rm rot})$. Since the
transformations are length and orientation-preserving, the
transformation of the velocity field is also preserved as the velocity
field generated by the right-hand rotation about the axis
$(\Theta_{\rm rot},\Phi_{\rm rot})$.  The radial velocity function is
simply the $z$-component of this velocity field in the dynamic
coordinate system. Namely,
\begin{align}
\begin{aligned}\label{eq:v_xyomega}
v(x,y)=\omega \bigg[\bigg(\sin(\Phi_{\rm rot})\cos(\Theta_{\rm rot})\bigg)x+
\bigg(\sin(\Phi_{\rm rot})\sin(\Theta_{\rm rot})\bigg)y\bigg]
\end{aligned}
\end{align}
for $(x,y)$ being any point on the POS disk normalized to the unit disk. 
Define:
\begin{align}
\begin{aligned}\label{eq:AB_definition}
& A \equiv \sin(\Phi_{\rm rot})\cos(\Theta_{\rm rot}) \\
& B \equiv \sin(\Phi_{\rm rot})\sin(\Theta_{\rm rot})
\end{aligned}
\end{align}
Equation 27 in ${\rm Paper~I}$ models the radial velocity perturbation (R-M effect) during 
an eclipse of the star as follows:
\begin{equation}\label{eq:deltaRV2}
\delta RV = \frac{\mathlarger{\iint}\displaylimits_{D_{\rm vis}}(Ax+By)\cdot IdA }{\mathlarger{\iint}\displaylimits_{D_{\rm{vis}}}IdA} =
\frac{\mathlarger{\oint}\displaylimits_{\partial D_{\rm vis}}\big[P_{\rm RM},Q_{\rm RM}\big] \boldsymbol{\cdot} \big[x', y'\big] d\varphi}{\mathlarger{\oint}\displaylimits_{\partial D_{\rm{vis}}}\big[P_{\rm FF},Q_{\rm FF}\big] \boldsymbol{\cdot} \big[x', y'\big] d\varphi}                
 \end{equation}
The numerator is the rotational velocity field, moderated by the intensity of the star, $I$. 
The denominator is the normalized intensity. From 
Equation (\ref{eq:RMgfunc_summary_appendix}) we 
have 
\begin{align}
\begin{aligned}\label{eq:wedge_intensity}
&d\wedge\bigg[0,G(r^2)\bigg]=xI(r^2)  \\
&\text{and} \\
&d\wedge\bigg[-G(r^2),0\bigg]=yI(r^2)
\end{aligned}
\end{align}
Thus,
\begin{align}
\begin{aligned}\label{eq:wedge_AB}
&d\wedge\bigg[-BG(r^2),AG(r^2)\bigg]=\big(Ax+By)I(r^2)  \\
\end{aligned}
\end{align}
Hence
\begin{equation}\label{eq:deltaRV2_BA}
\delta RV = \frac{\mathlarger{\oint}\displaylimits_{D_{\rm vis}}
G(r^2)\big(-Bx^\prime+Ay^\prime\big)d\varphi}
{\mathlarger{\oint}\displaylimits_{D_{\rm vis}}
\frac{G(r^2)}{r^2}\big(-yx^\prime+xy^\prime\big)d\varphi}
\end{equation}
Applying this method to the Hot Jupiter example with the additional
parameters: $i=89^\circ$, rotation period $=20 {\rm~days}$,
$\Phi_{\rm rot}=90^\circ$ (colatitude), and $\Theta_{\rm rot}=0$
(angle from the POS $y$ axis), we obtain the result shown in Figure
\ref{fig:RMEffect90_0HotJupIP25}. Note that ``$N$'' is the north pole
of the right handed stellar rotation axis. The black circle is the
outline of the planet, and the black line is its orbital track. If we
now change the axis of the stellar rotation to $\Phi_{\rm
  rot}=30^\circ$ (colatitude), and $\Theta_{\rm rot}=-80$ (angle from
the POS $y$ axis), we obtain what is shown in Figure
\ref{fig:RMEffect30_m80HotJupIP25}.  \\

\section{Table Defined Functions}\label{sec:appendix_tabledefined}

Given a table of values (e.g \. an Atmosphere Intensity table)
\begin{align}
\begin{aligned}\label{eq:table_of_values}
\begin{bmatrix}
 r_1^2  & r_2^2  & \boldsymbol{\ldots} & r_n^2 \\
 I_1 & I_2 & \boldsymbol{\ldots} & I_n \\
\end{bmatrix}
\end{aligned}
\end{align}
a piecewise linear function $f_{\rm PL}$ is defined by 
\begin{align} \label{eq:f_pl_define}
\begin{aligned}
I_i=f_{\rm PL}(r_i^2)\end{aligned}
\end{align}
For intermediate values the intensities are found by linear interpolation. 
For $r_i^2 < r^2 < r_{i+1}^2$  linear interpolation gives 
\begin{align} \label{eq:fpl_linearinterp}
\begin{aligned}
&f_{\rm PL}(r^2) = I_i+a(r^2 - r_i^2) \\
&{\rm where~~}
a=\frac{I_{i+1}-I_i}{r_{i+1}^2-r_i^2}
\end{aligned}
\end{align}
Equation (\ref{eq:RMgfunc_integral_appendix}) defines an important integration
operation on $f_{\rm PL}$, namely
\begin{align}\label{eq:RMgfunc_integral_fpl}
\begin{aligned}
G(r^2)=\frac{1}{2}\int\displaylimits_0^{r^2}I(z)dz \\ 
\end{aligned}
\end{align}
To build the table of values we assume $r_1^2=0$, and for subsequent table values 
we obtain a recursion formula starting with $G(r_1^2)=0$ and continuing with 
\begin{align}\label{eq:recurrsion_integ1}
\begin{aligned}
G(r_j^2)=\frac{1}{2}\int\displaylimits_0^{r_j^2}f_{\rm PL}(z)dz = 
\frac{1}{2}\int\displaylimits_0^{r_{j-1}^2}f_{\rm PL}(z)dz +
\frac{1}{2}\int\displaylimits_{r_{j-1}^2}^{r_j^2}f_{\rm PL}(z)dz
\end{aligned}
\end{align}
or
\begin{align}\label{eq:recurrsion_integ2}
\begin{aligned}
G(r_j^2)=G(r_{j-1}^2)+\frac{1}{2}\int\displaylimits_{r_{j-1}^2}^{r_j^2}f_{\rm PL}(z)dz
\end{aligned}
\end{align}
where $f_{\rm PL}(z)$ is the line segment from $(r_{j-1}^2,I_{j-1})$ to $(r_{j}^2,I_{j})$.
The value of this integral is given exactly by the mid-point rule (Gaussian 1-point) as
\begin{align}\label{eq:integvalue_midptrule}
\begin{aligned}
\frac{1}{2}\int\displaylimits_{r_{j-1}^2}^{r_j^2}f_{\rm PL}(z)dz=
\frac{1}{2}\frac{ f_{\rm PL}(r_{j-1}^2)+f_{\rm PL}(r_{j}^2) }{2}(r_j^2-r_{j-1}^2)
=\frac{1}{4}(I_{j-1}+I_j)(r_j^2-r_{j-1}^2)
\end{aligned}
\end{align}
In addition, we can compute the derivative of $G$ with respect to $r^2$:
\begin{align}\label{eq:G_deriv}
\begin{aligned}
& & \frac{d}{dr^2}G(r^2)=\frac{1}{2}f(r^2)\\
\end{aligned}
\end{align}
or for the table values,
\begin{align}\nonumber
\begin{aligned}
& &\frac{d}{dr^2}G(r^2) \bigg|_{r_i^2}=\frac{1}{2}f(r_i^2)
\end{aligned}
\end{align}
Thus we obtain a new table based on the function $G(r^2)$:
\begin{align}
\begin{aligned}\label{eq:new_table_of_values}
\begin{bmatrix}
r_1^2 & r_2^2 &\boldsymbol{\ldots} & r_n^2 \\
G(r_1^2) & G(r_2^2) & \boldsymbol{\ldots} & G(r_n^2) \\
\frac{1}{2}I_1 & \frac{1}{2}I_2 &  \boldsymbol{\ldots} & \frac{1}{2}I_n
\end{bmatrix}
\end{aligned}
\end{align}
where the third row is the derivative of the second row with respect
to the first row. This table uniquely defines a piecewise cubic Hermit
(PCH) function using cubic Hermite interpolation for intermediate
values. That is, $G_{\rm PCH}(r_i^2)=G(r_i^2)$, and for $r_i^2 < r^2 <
r_{i+1} ^2 ~~~ G_{\rm PCH}(r^2)=\text{cubic Hermite Interpolant}$. \\ \\
\begin{align}
\begin{aligned}
\setlength\extrarowheight{-15pt}
\begin{tabular}{c|lccc}
\centering
$r_i^2$ &  $G(r_i^2)$  &                                  &          &  \\
             &                     &  $G^\prime(r_i^2)$   &          &    \\
$r_i^2$ &   $G(r_i^2)$ &                                  &  $~~b$  &        \\
            &                     &  $a$                         &           &  $~~~~~~d$        \\
$r_{i+1}^2$  &   $G(r_{i+1}^2)$  &                   &        $~~c$      &           \\
  \\          &      &                 $G^\prime(r_{i+1}^2)$        &       &           \\
$r_{i+1}^2$            &   $G(r_{i+1}^2)$                     &       &       &           \\
\end{tabular} \\ \\ \\
\end{aligned}
\end{align}
where
\begin{align}
\begin{aligned}\nonumber
a=\frac{G(r_{i+1}^2)-G(r_i^2)}{r_{i+1}^2-r_i^2}~,~
b=\frac{a-G^\prime(r_i^2)}{r_{i+1}^2-r_i^2}~,~
c=\frac{G^\prime(r_{i+1}^2)-a}{r_{i+1}^2-r_i^2}~,~
d=\frac{c-b}{r_{i+1}^2-r_i^2}
\end{aligned}
\end{align}
From this divided difference table, the cubic interpolating polynomial is given by:
\begin{align}
\begin{aligned}\label{eq:G_PCH}
G_{\rm PCH}(r^2)=G(r_i^2)+(r^2-r_i^2)
\bigg\{G^\prime(r_i^2)+(r^2-r_i^2)\bigg[b+d(r^2-r_{i+1}^2)\bigg]\bigg\}
\end{aligned}
\end{align}
From Equation (\ref{eq:RMgfunc_summary_appendix}) we also
need to construct the PCH function $G_{\rm PCH}(r^2)/r^2$. To do this, 
we must compute
its derivative
\begin{align}
\begin{aligned}\label{eq:derivative_eq87}
&\frac{d}{dr^2}\bigg(G(r^2)/r^2\bigg)=\frac{G^\prime(r^2)}{r^2}-\frac{G(r^2)}{r^4} \\
&\text{and} \\
&\lim_{r^2 \to  0} \frac{G(r^2)}{r^2},~~\lim_{r^2 \to 0}\frac{d}{dr^2}\bigg(G(r^2)/r^2\bigg) \\
\end{aligned}
\end{align}
The first limit is evaluated by L'Hospital's Rule:
\begin{align}
\begin{aligned}\label{eq:derivative_eq87_firstlimit}
&\lim_{r^2 \to  0} \frac{G(r^2)}{r^2}=&\lim_{r^2 \to  0} \frac{G^\prime(r^2)}{1}=\frac{1}{2}I(0) \\
\end{aligned}
\end{align}
to compute the second limit, assume that the intensity is a maximum at
$r^2=0,$\footnote{see our note on averaging intensities at the end of this section} and that the intensity is smooth at $r^2=0$. Expand the
intensity about $r^2=0$ in a Taylor series:
\begin{align}
\begin{aligned}\label{eq:intensity_taylor}
I(r^2)&=I(0)+I^\prime(0)r^2+O(r^4) \\ \\
 \text{and}  \\
G(r^2)&=\frac{1}{2}\int\displaylimits_0^{r^2}I(0)+I^\prime(0)z+O(z^2)dz \\
&=\frac{1}{2}\bigg(I(0)r^2+I^\prime(0)\frac{r^4}{2} + O(r^6)\bigg)
\end{aligned}
\end{align}
\begin{align}
\begin{aligned}\label{eq:intensity_taylor_limit}
\lim_{r^2 \to 0} \frac{G^\prime(r^2)}{r^2}-\frac{G(r^2)}{r^4}=
\lim_{r^2 \to 0} \frac{1}{2} \frac{I(r^2)}{r^2}-\frac{1}{2}\bigg(\frac{I(0)}{r^2}+\frac{I^\prime(0)}{2}+O(r^2)\bigg)=
\frac{-I^\prime(0)}{4}
\end{aligned}
\end{align}
however, $I$ assumes a maximum at $r^2=0$, implying that $I^\prime(0)=0$. Thus the
table representation of $G(r^2)/r^2$ is given by:
\begin{align}
\begin{aligned}\label{eq:new_new_table_of_values}
\begin{bmatrix}
r_1^2 & r_2^2 &\boldsymbol{\ldots} & r_n^2 \\
\frac{1}{2}I(0) & \frac{G(r_2^2)}{r_2^2} & \boldsymbol{\ldots} & \frac{G(r_n^2)}{r_n^2}  \\
0 & \frac{\frac{1}{2}I_2}{r_2^2}-\frac{G(r_2^2)}{r_2^4} & \boldsymbol{\ldots} & 
       \frac{\frac{1}{2}I_n}{r_n^2}-\frac{G(r_n^2)}{r_n^4}
\end{bmatrix}
\end{aligned}
\end{align}
for efficiency, we include two additional table-rows in the code, 
the interpolation factors $b$ and $d$.  This will allow direct use of the cubic
interpolating polynomial without computing the difference table. \\

As a check, we will now evaluate the exterior 
derivative in Equation 87 of ${\rm Paper~I}$:
\begin{align}\label{eq:wedge_evaluate_eq87}
\begin{aligned}
&d\wedge\bigg[-y\frac{G(r^2)}{r^2},x\frac{G(r^2)}{r^2}\bigg]=I(r^2) \\
\text{the left hand side:} \\
&=\frac{\partial}{\partial x} x\frac{G(r^2)}{r^2}- 
\frac{\partial}{\partial y} y\frac{G(r^2)}{r^2} \\
&=\frac{G(r^2)}{r^2}+2x^2\frac{d}{dr^2}\frac{G(r^2)}{r^2}
+\frac{G(r^2)}{r^2}+2y^2\frac{d}{dr^2}\frac{G(r^2)}{r^2} \\
&=2\frac{G(r^2)}{r^2}+2r^2\frac{d}{dr^2}\frac{G(r^2)}{r^2} = I(r^2)
\end{aligned}
\end{align} \\
or in terms of the table representation of $G(r^2)/r^2$ we have 
\begin{align}
\begin{aligned}
2\cdot {\rm row2 } +  2\cdot {\rm row1} \ast {\rm row3}=
{\rm Intensity} \nonumber
\end{aligned}
\end{align}
where $\ast$ is element 
by element multiplication. The result is that  the manulipation
of the table rows in the above manner gives the original intensity
to the machine precision. \\

Finally, we comment on a lesser known issue of modeling eclipses with limb darkening laws where the boundary of the front body crosses the back body's center.  The limb-darkening laws are not models of stellar intensity, but rather are  models of {\it time and azimuthal averaged} stellar intensities. This averaging eliminates the time varying processes such as spots, granulation, etc.  The averaging produces a smooth POS stellar intensity model with no angular dependence.  Thus, the derivative of this two dimensional function exists at the star's center and is angular-independent there.  It then follows that the partial derivatives with respect to x and y exist at the center and must be zero, implying that the center is a critical point. Because the stellar intensity increases as one moves towards the center, the center must be a maximum for the two dimensional stellar intensity function. Thus for its one dimensional cross-section the derivative at $\mu=1$ (the center) is zero.
In contrast, for the standard limb darkening laws the one-sided derivative at the center is not zero.
This implies that the limb darkening model applied to the POS is not differentiable at the center.
If we are using the Mandel-Agol algorithm to model an eclipse where the center of the back body is covered, at least two different special function approximations are required (this is the case of crossing from ``region 2'' to ``region 8'' or crossing from ``region 3'' to ``region 9'', see Figure 6 in ${\rm Paper~I}$). 
This may result in an error in the computation of the transit profile (see Figure 17 in ${\rm Paper~I}$).
Note that if a numerical integration scheme is used instead of a set of special functions,
the discontinuity at the center does not lead to errors in the flux. That numerical scheme may use tabulated stellar intensities (this paper) or the limb darkening laws themselves (e.g.\ ${\rm Paper~I}$). It is the use of the special functions and the matching of boundary conditions that leads to the error in the transit profile.

\section{Error Analysis for the Atmosphere Table 
Method}\label{sec:atm_err_analysis}

Suppose we have an error estimate for the entries of a stellar
atmosphere table. We can ask two questions about the error associated
with that table. The first question is, what is the resulting maximum
error for the light curve of an eclipse computed from such a table?
The second question is, given an atmosphere table of size $N_\mu$, what
is the maximum absolute error generated by using the table PL function
in the computation of a light curve?\\

To address the first question, 
assume that the table center-to-limb intensities are given by
\begin{align} \label{eq:center_limb_int}
\begin{aligned}
I(\mu_i)=I_{\rm AT}(\mu_i)+E(\mu_i)
\end{aligned}
\end{align}
where $E$ is an error distribution of intensities, having a mean 0 and
a variance $\sigma_{E}^2$. Further assume that the Central Limit
Theorem applies for $N \ll N_\mu$ and that the mean of
the sum of $N_\mu$ error
distributions closely approximates a normal distribution with a mean 0
and variance $\sigma_{E}^2/N_\mu$. For any selection of errors using
the prescribed distribution, $I$ is just a new stellar atmosphere
table, and as such we may apply the methodology of Appendix
\ref{sec:appendix_tabledefined}. First, the independent variable $\mu$
is changed to $r^2$.  Equation \ref{eq:RMgfunc_integral_fpl} shows
that the function $G(r^2)$ is linear in $I$, which implies
\begin{align} \label{eq:Gofr2_linear_implies}
\begin{aligned}
&G_I(r^2)=G_{\rm AT}(r^2)+G_E(r^2) \\
&\frac{G_I(r^2)}{r^2}=\frac{G_{\rm AT}(r^2)}{r^2}+\frac{G_E(r^2)}{r^2}
\end{aligned}
\end{align}

Finally, computing the flux (Equations \ref{eq:greenapply},
\ref{eq:limbdarkintegral_new1}, and \ref{eq:limbdarkintegral_new2}) we
note that flux is linear with respect to the function $G/r^2$, thus we
obtain for the Flux $\cal{F}$:
\begin{align} \label{eq:flux_pz}
\begin{aligned}
{\cal{F}}_I(p,z)={\cal{F}}_{\rm AT}(p,z)+{\cal{F}}_E(p,z)
\end{aligned}
\end{align}
where $(p,z)$ describes the geometry of the eclipsing system as in
Section \ref{sec:implementation} and ${\rm Paper~I}$. The flux is then
normalized by the flux of the unobstructed body, which is given by:
\begin{align} \label{eq:flux_unobstructed_body}
\begin{aligned}
{\cal{F}}_{\rm out~of~eclipse}={\cal{F}}_0=\frac{2 \pi G_{\rm AT}(1)}{1} 
 + \frac{2 \pi G_E(1)}{1}
\end{aligned}
\end{align}
Our immediate goal is to reduce $G_E(1)$ to a small fraction of
$G_{\rm AT}(1)$, so that ${\cal{F}}_0$ is simply $2 \pi G_{\rm
  AT}(1)$. Applying the trapezoid rule to $G_E(1)$ (which is exact
since E(z) is a PL function), we obtain
\begin{align} \label{eq:GE1_trapezoid}
\begin{aligned}
G_E(1)=\frac{1}{2}\int\displaylimits_{0}^{1}E(z)dz
=\frac{\frac{1}{2}\bigg\{\bigg[\sum\displaylimits_{i=2}^{N_\mu-1}E(\mu_i)\bigg]+
\frac{1}{2}\bigg[E(0)+E(1)\bigg]\bigg\}}{N_\mu -1}
\end{aligned}
\end{align}
Since $I(1)=1$ then for all atmosphere tables $E(1)=0$:
\begin{align} \label{eq:GE1_trapezoid2}
\begin{aligned}
&G_E(1)=\frac{\frac{1}{2}\bigg\{\bigg[\sum\displaylimits_{i=2}^{N_\mu-1}E(\mu_i)\bigg]+
\frac{1}{2}\bigg[E(0)\bigg]\bigg\}}{N_\mu -1}
\end{aligned}
\end{align}
or
\begin{align} \label{eq:GE1_distribution}
\begin{aligned}
{\rm Distribution(G_E(1))=\frac{1}{2}~Distribution(mean(E+E...(N_\mu-1)~times))}
\end{aligned}
\end{align}
From our beginning assumption, the Central Limit Theorem implies that
the mean of the sum of the error distributions is normal with mean 0
and variance $\sigma_E^2/(N_\mu-1)$. While this normal distribution is 
unbounded, large values are rare. For example, if we cut off the tail of the 
distribution by 5 sigmas, assume a table size of 1001, and $\sigma_E=0.01$, 
and noting that $0.37 < G_{\rm AT}(1) < 0.43$, for all of the models in the Kepler band 
produced by \citet{Neilson2019}, we obtain:
\begin{align} \label{eq:GE1_unbound}
\begin{aligned}
&\mid G_E(1)\mid<\frac{1}{2}\bigg(\frac{5\sigma_E}{\sqrt{N_\mu-1}}\bigg) 
\frac{G_{\rm AT}(1)}{0.37}=0.0021G_{\rm AT}(1) \\
\end{aligned}
\end{align}
Thus, as the table size grows, the standard deviation of the error
is reduced and the contribution of the error to the
out-of-eclipse flux becomes more insignificant.
\begin{align} \label{eq:Eq_modified_unitcirc}
\begin{aligned}
& {\cal{F}}_0=2 \pi G_{\rm AT}(1) \text{~~~Equation 
\ref{eq:limbdarkintegral_new2} with the arc being the unit circle} \\ 
& \Rightarrow \frac{ {\cal{F}}_I(p,z) }{ {\cal{F}}_0 }=
\frac{ {\cal{F}}_{\rm AT}(p,z) }{ {\cal{F}}_0 }+\frac{ {\cal{F}}_E(p,z) }{ {\cal{F}}_0 } \\
& \Rightarrow \text{MaxLightCurve Error}={\rm max_{(p~and~z)}}\bigg\lvert\frac{ {\cal{F}}_E(p,z) }{ {\cal{F}}_0 } \bigg\rvert
\end{aligned}
\end{align}
for the region in the (p,z) plane given by $0 < p < 3$ and $0 < z <
3$. 
We will compute the maximum light curve error for two
cases: \\

{\it Case 1}: $E$ is the normal distribution with mean 0 and standard
deviation $\sigma$. $E$ can be written as $E=\sigma U$, where $U$ is
the unit normal distribution having mean 0 and standard deviation 1.
\begin{align} \label{eq:MaxLCError_case1}
\begin{aligned}
&\text{MaxLightCurveError}={\rm max_{(p~and~z)}} \bigg\lvert 
\frac{ \sigma{\cal{F}}_{\rm U}(p,z) }{ {\cal{F}}_0 } \bigg\rvert =
& \frac{\sigma}{2 \pi G_{\rm AT}(1)}{\rm max_{(p~and~z)}} \bigg\lvert 
{\cal{F}}_{\rm U}(p,z) \bigg\rvert
\end{aligned}
\end{align}
Only the first term, ${\sigma}/ [2 \pi G_{\rm AT}(1)]$, depends on
the particular atmosphere table and its size. For the second term,
10,000 tables were constructed for the table size $N_\mu=1001$. For
each of these, the ${\rm max_{(p~and~z)}}\lvert {\cal{F}}_{\rm
  U}(p,z) \rvert$ was computed. Figure
\ref{fig:DistriMaxErrorUnitNormal} shows a histogram of the resulting
light curve maxima. For these 10,000 sample tables of the unit normal
distribution, the maximum absolute value error was 0.1748. Since the
error modeled by the unit normal distribution can have arbitrary large
values - even if very rare - the actual maximum does not
exist. Truncating the normal distribution at a few $\sigma$ would
solve this problem. We then have
\begin{align} \label{eq:MaxLCError_case1_num}
\begin{aligned}
\text{MaxLightCurveError}=\frac{\sigma}{2 \pi 
G_{\rm AT}(1)}{\rm max_{(p~and~z)}} \bigg\lvert {\cal{F}}_{\rm U}(p,z) \bigg\rvert
=\frac{\sigma}{2 \pi G_{\rm AT}(1)}0.18=\frac{0.18}{{\cal{F}}_0}\sigma
\end{aligned}
\end{align}
For the set of 918 atmosphere tables by \citet{Neilson2013} in the
$Kepler$ band, the bounds on ${\cal{F}}_0$ are $2.330 < {\cal{F}}_0 <
2.652$, giving the result: $\text{\it
  Kepler~}\text{MaxLightCurveError}=0.077\sigma$ where $\sigma$ is the
standard deviation of the original table error. \\

{\it Case 2}: $E$ is the uniform distribution on $[-0.5,0.5]\epsilon$
with mean 0 and variance ${\epsilon^2}/{12}$. This case arises
when we have a fixed number of decimal digits.  Applying the Central
Limit Theorem, the ${\rm Distribution}(G_E(1))$ is a normal
distribution with mean 0 and variance
${(N_\mu-1)\epsilon^2}/{24}$ (see Equation
\ref{eq:GE1_unbound}). Doing the same computation as done in {\it
  Case 1}, we obtain the distribution of the maximum absolute error in
the light curve computation. Figure \ref{fig:DistriMaxErrorUnitUniform} shows a histogram of the resulting
light curve maxima.  For the same $Kepler$ band \citet{Neilson2013}
atmosphere tables we obtain: $\text{\it Kepler~}\text{MaxLightCurveError}=0.022\epsilon$, 
where ${\epsilon^2}/{12}$ is the variance of the original table
error. Note that the ratio of the Normal Error estimate and the
Uniform Error estimate is ${1}/{\sqrt(12)}\approx 0.289$. \\

We now address the second question raised at the beginning of
this section.
Since the error induced by PL
approximations is $O(\Delta x^2)$, one would expect doubling the table
size would reduce the error by a factor of 4. The approach to
answering this question is to construct a much better table
approximation than provided by a PL function. If we use the central
difference approximation to the derivative of the intensity at each
interior value of $\mu_i$, we obtain a second order approximation to
$dI_i$. Thus we have the table $\big [\mu_i, I_i, dI_i\big]$ which
defines a PCH function approximation of $I$. Rather than connecting
the intensity values with line segments, where each adjacent pair is
now connected by a cubic polynomial such that the PCH function is a
continuous differentiable function for all values of $\mu$. For each
of the 918 \citet{Neilson2013} tables using the $Kepler$ band, we
compute the maximum light curve difference between the PL and the PCH
function approximations for the region in the $(p,z)$ plane given by
$0 < p < 3$ and $0 < z < 3$. 
Figure \ref{fig:MaxLCErrorforNandLTables} shows the maximum
light curve error generated by the PL function defined by a table size
$N_\mu=1001$ for all of the 918
models in the \citet{Neilson2013} grid.  The maximum error estimate
is  $2.1\times10^{-7}$.  The mean and median values from the grid
are $7.15\times 10^{-8}$ and $6.79 \times 10^{-8}$, respectively.
Figure
\ref{fig:Error10kminus1000t5700g450m11} shows a specific example,
where we consider the \citet{Neilson2013} table for $T_{\rm
  eff}=5700,~\log(g)=4.50$, and mass $M=1.1M_\odot$.

\acknowledgments
This material is based upon work supported by the National Science Foundation under Grant No. (NSF AST-1617004). We are
also deeply grateful to John Hood, Jr. for his generous support of exoplanet research at San Diego State University.


\begin{deluxetable}{cccc}[t]
\tablenum{1}
\tablecaption{Error Statistics}\label{tab:moderror}
\tablewidth{0pt}
\tablehead{
 $T$    & $N_{\mu}$  & maximum & median  \\
        &            & difference & difference 
}
\startdata
1   &  1001  & $2.143\times 10^{-6}$    &  $4.510\times 10^{-7}$ \\
2   &  1001  & $3.806\times 10^{-7}$    &  $4.248\times 10^{-8}$ \\
4   &  1001  & $7.356\times 10^{-8}$    &  $1.443\times 10^{-8}$ \\
8   &  1001  & $7.356\times 10^{-8}$    &  $1.209\times 10^{-8}$ \\
1   &  2001  & $2.145\times 10^{-6}$    &  $4.434\times 10^{-7}$ \\
2   &  2001  & $3.837\times 10^{-7}$    &  $2.968\times 10^{-8}$ \\
4   &  2001  & $6.762\times 10^{-8}$    &  $4.817\times 10^{-9}$ \\
8   &  2001  & $1.839\times 10^{-8}$    &  $3.089\times 10^{-9}$ \\
1   &  4001  & $2.145\times 10^{-6}$    &  $4.415\times 10^{-7}$ \\
2   &  4001  & $3.845\times 10^{-7}$    &  $2.709\times 10^{-8}$ \\
4   &  4001  & $6.803\times 10^{-8}$    &  $2.001\times 10^{-9}$ \\
8   &  4001  & $1.205\times 10^{-8}$    &  $8.694\times 10^{-10}$ \\
1   &  8001  & $2.145\times 10^{-6}$    &  $4.410\times 10^{-7}$ \\
2   &  8001  & $3.846\times 10^{-7}$    &  $2.645\times 10^{-8}$ \\
4   &  8001  & $6.813\times 10^{-8}$    &  $1.132\times 10^{-9}$ \\
8   &  8001  & $1.023\times 10^{-8}$    &  $3.019\times 10^{-10}$ \\
\enddata
\end{deluxetable}


\begin{figure}[ht!]\
\includegraphics[angle=90,scale=0.7]{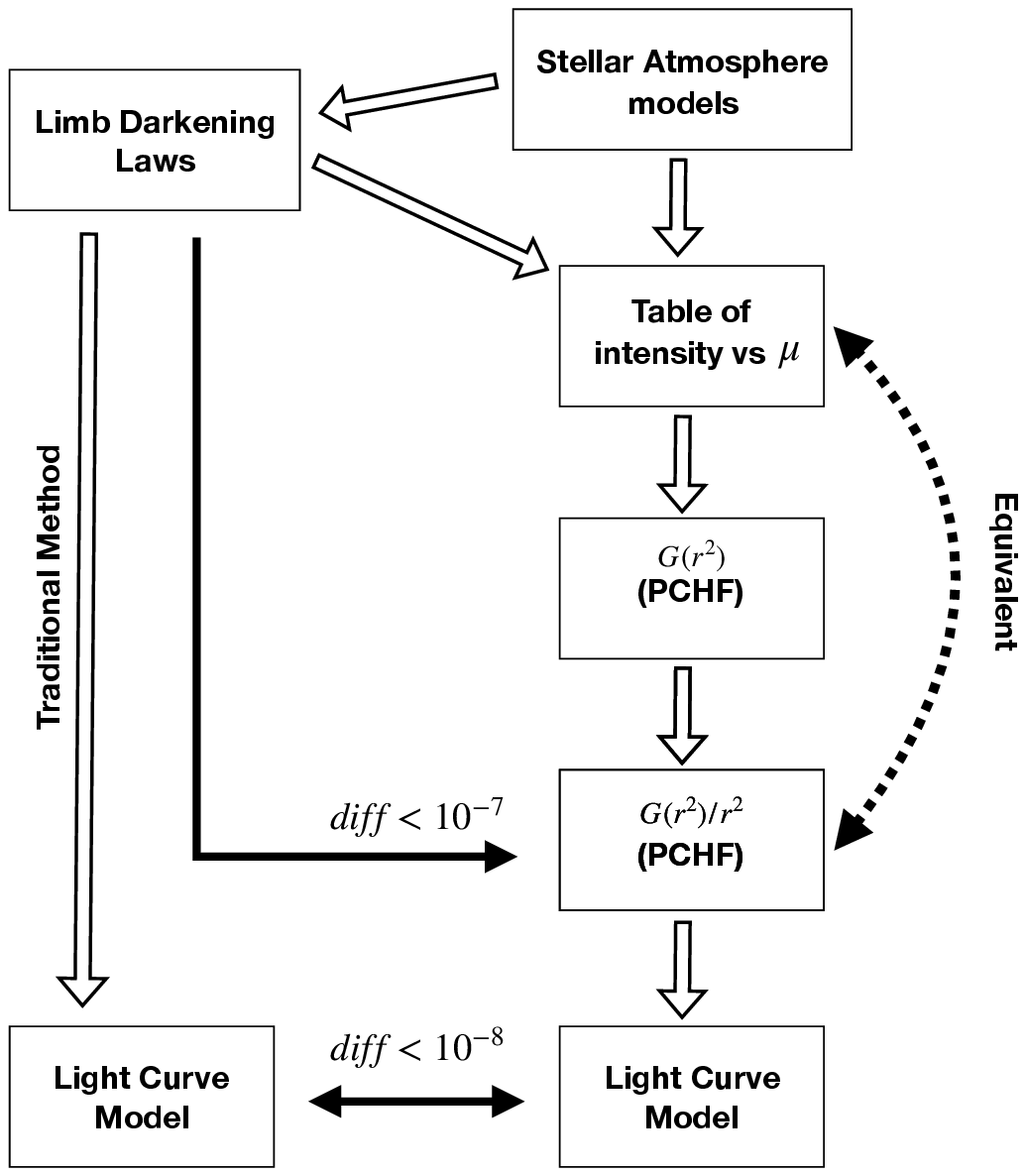}
\caption{An overview of the method described in this work. The
starting point is a Table of $\mu$ with the corresponding intensity
values. Such a table may be generated by a stellar atmosphere model
or by a parametrized analytic limb darkening law. This $(\mu,I)$
table then generates a piecewise cubic Hermite function (PCHF),
$G(r^2)$, from Equation \ref{eq:RMgfunc_integral}, and another PCHF
$G(r^2)/r^2$. These PCHFs are created without error. The dotted
arrow going back to the original table, indicates that the original
table can be recreated without loss from the PCHF $G(r^2)/r^2$. 
Equation \ref{eq:greenapply} gives the path integrals and equation
\ref{eq:RMgfunc_summary} gives the 1-form $\big[P, Q\big]$. Hence,
the PCHF $G(r^2)/r^2$ is the desired form for the atmosphere table. 
This form directly generates the integrands of the path
integral. The dark arrows indicate comparisons with the standard
limb darkening laws assuming a table size of 1001 equally spaced
values of $\mu$ in $[0, 1]$.}
\label{fig:method_overview}
\end{figure}

\begin{figure}[ht!]\
\plotone{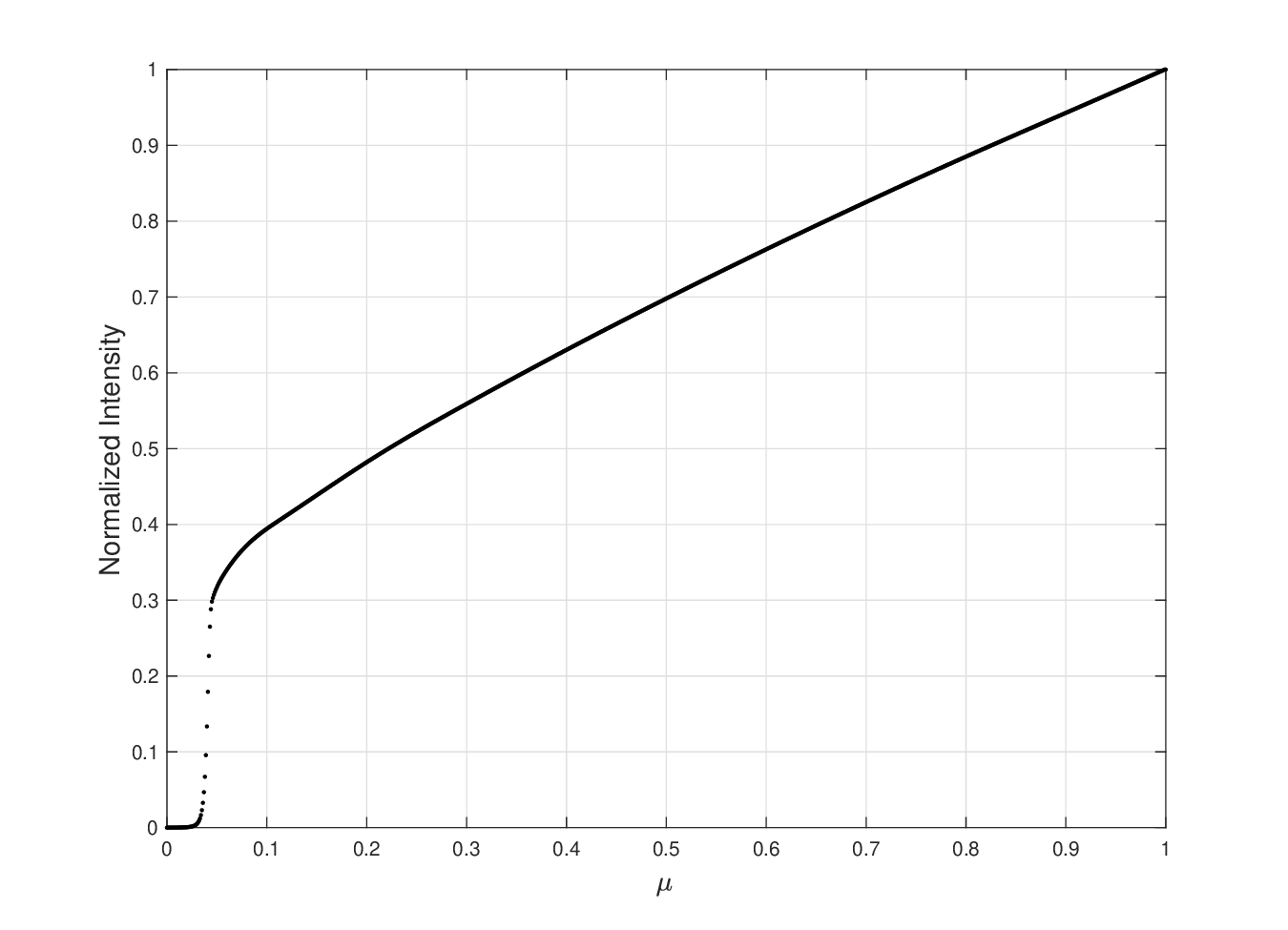}
\caption{The normalized intensity vs $\mu$ for the Spherical Model
  Stellar Atmosphere Table from \citet{Neilson2013} for the stellar
  parameters $T_{\rm eff} = 5700K$, $\log(g)=4.50$ and
  Mass=$1.1M_\odot$ in the $Kepler$ band.}
\label{fig:muVIntensity}
\end{figure}

\begin{figure}[ht!]\
\plotone{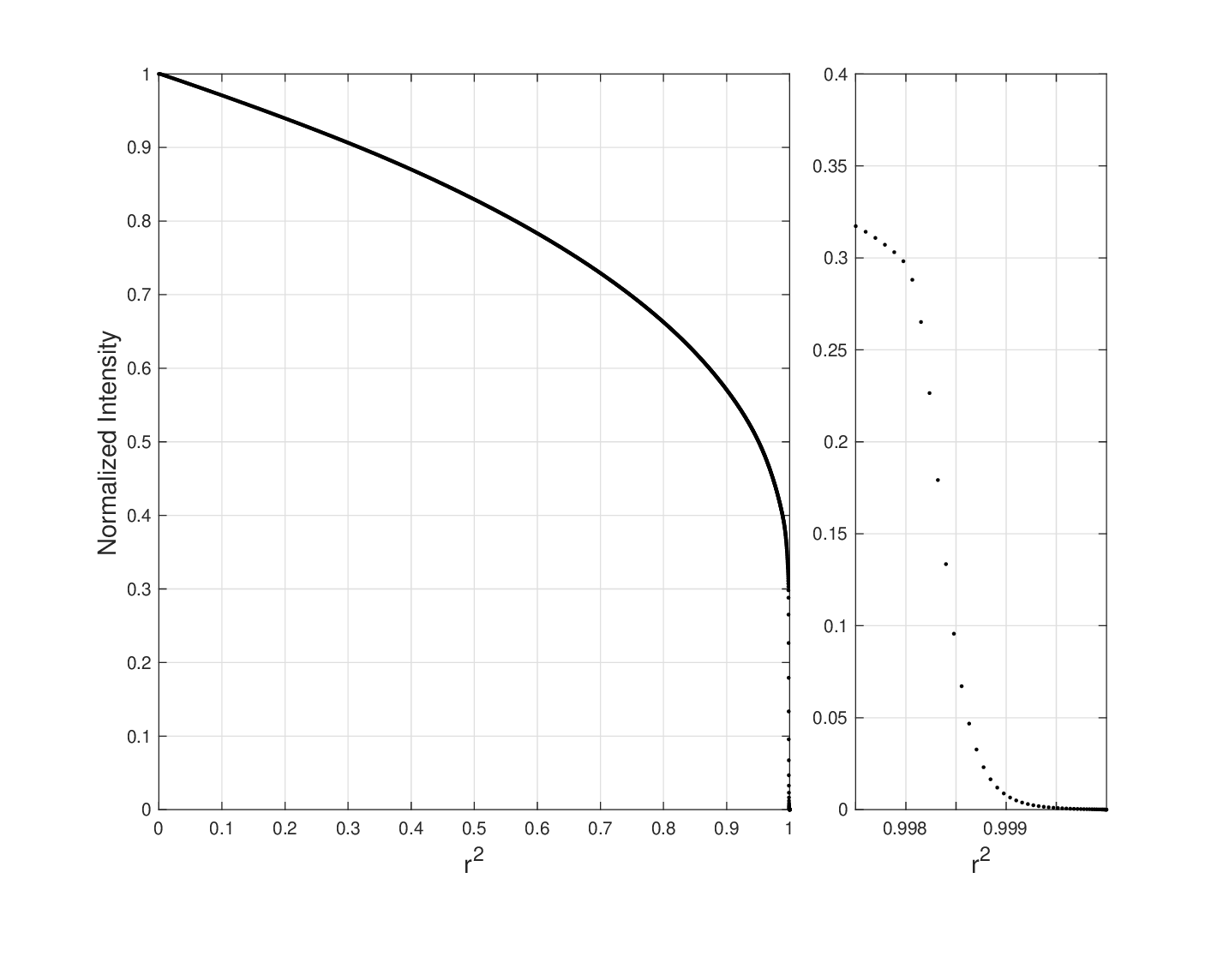}
\caption{The normalized intensity vs $r^2$ for the Spherical Model
  Stellar Atmosphere Table from \citet{Neilson2013} for the stellar
  parameters $T_{\rm eff} = 5700K$, $\log(g)=4.50$ and
  Mass=$1.1M_\odot$ in the $Kepler$ band.}
\label{fig:rsqVIntensity}
\end{figure}

\begin{figure}[ht!]\
\plotone{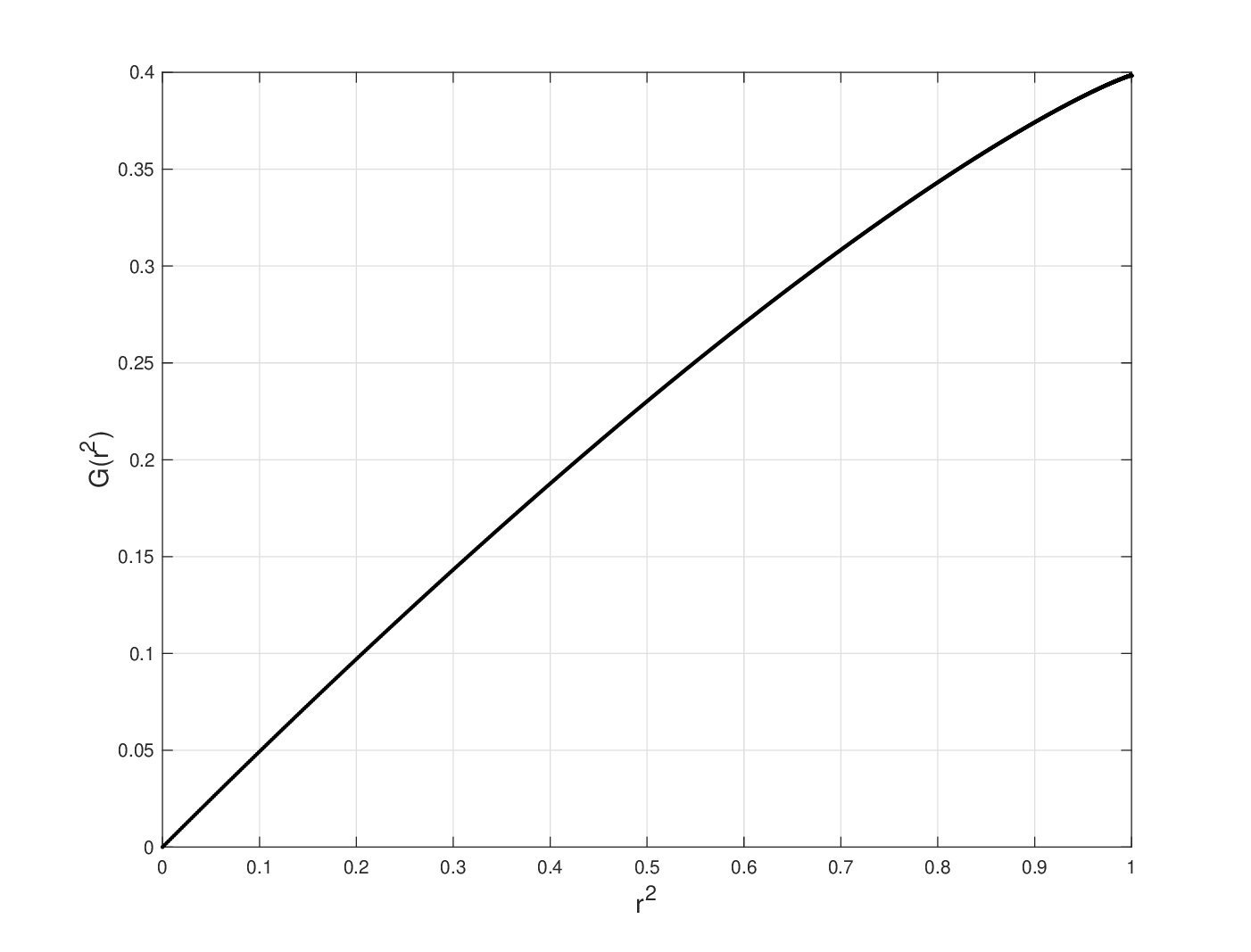}
\caption{$G(r^2)$ vs $r^2$ for the Spherical Model Stellar Atmosphere
  Table from \citet{Neilson2013} for the stellar parameters $T_{\rm
    eff} = 5700K$, $\log(g)=4.50$ and Mass=$1.1M_\odot$ in the
  $Kepler$ band.}
\label{fig:rsqVGrsq}
\end{figure}

\begin{figure}[ht!]\
\plotone{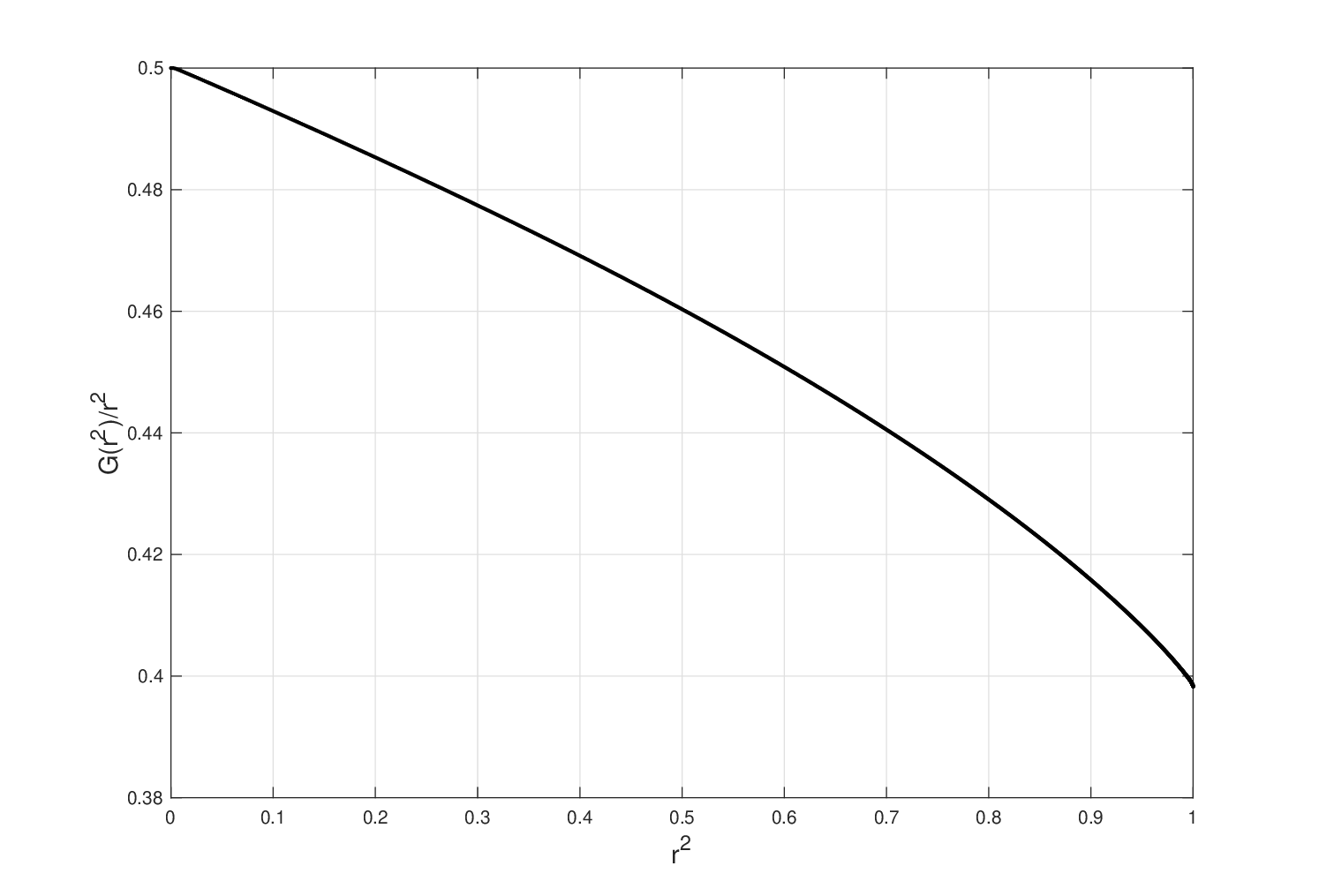}
\caption{$G(r^2)/r^2$ vs $r^2$ for the Spherical Model Stellar
  Atmosphere Table from \citet{Neilson2013} for the stellar parameters
  $T_{\rm eff} = 5700K$, $\log(g)=4.50$ and Mass=$1.1M_\odot$ in the
  $Kepler$ band.}
\label{fig:rsqVGrsqDbyrsq}
\end{figure}

\begin{figure}[ht!]\
\plotone{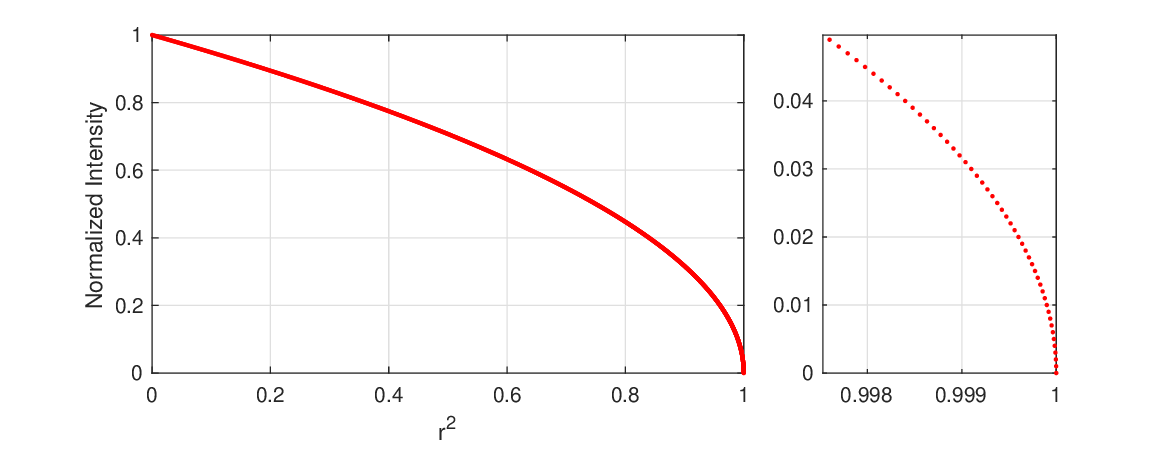}
\caption{Normalized intensity vs $r^2$ for the QL's Linear Basis
  Function $I=\mu(r^2)$.}
\label{fig:LinearBasisFrsq}
\end{figure}

\begin{figure}[ht!]\
\plotone{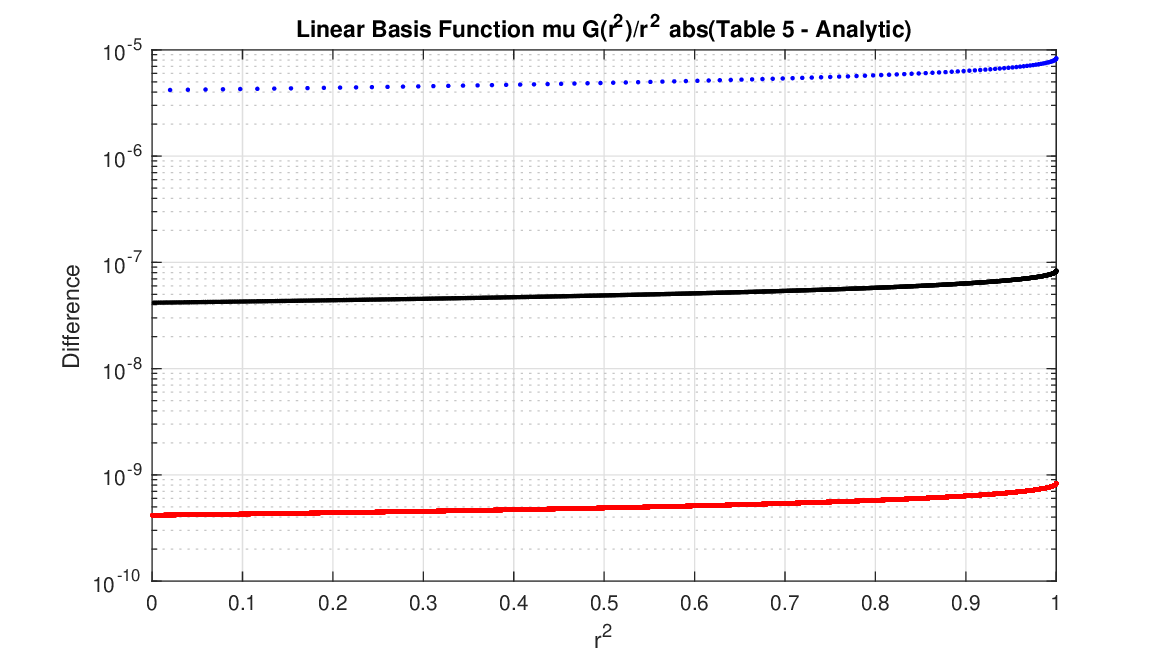}
\caption{The difference between the QL Linear Basis Function's
  analytic form for $G(r^2)/r^2$ and the Atmosphere form of $G_{\rm
    PCH}(r^2)/r^2$ .  The table sizes are denoted by blue=101,
  black=1001, and red=10001.}
\label{fig:LinearBasisGrsqDbyrsq}
\end{figure}	

\begin{figure}[ht!]\
\plotone{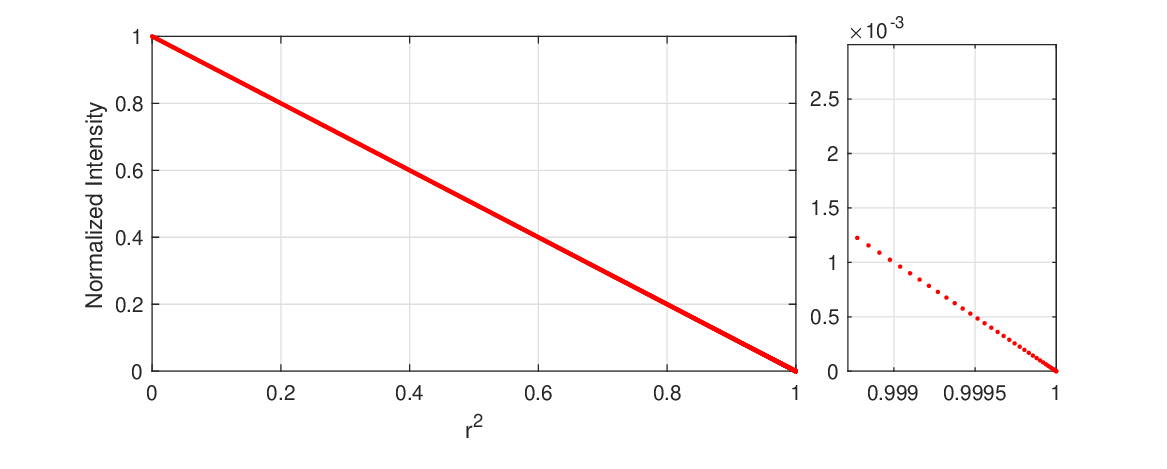}
\caption{Normalized intensity vs $r^2$ for the QL's Quadratic Basis function 
$I=\mu^2(r^2)$.}
\label{fig:QuadBasisFrsq}
\end{figure}

\begin{figure}[t]
\includegraphics[angle=-90,scale=0.65]{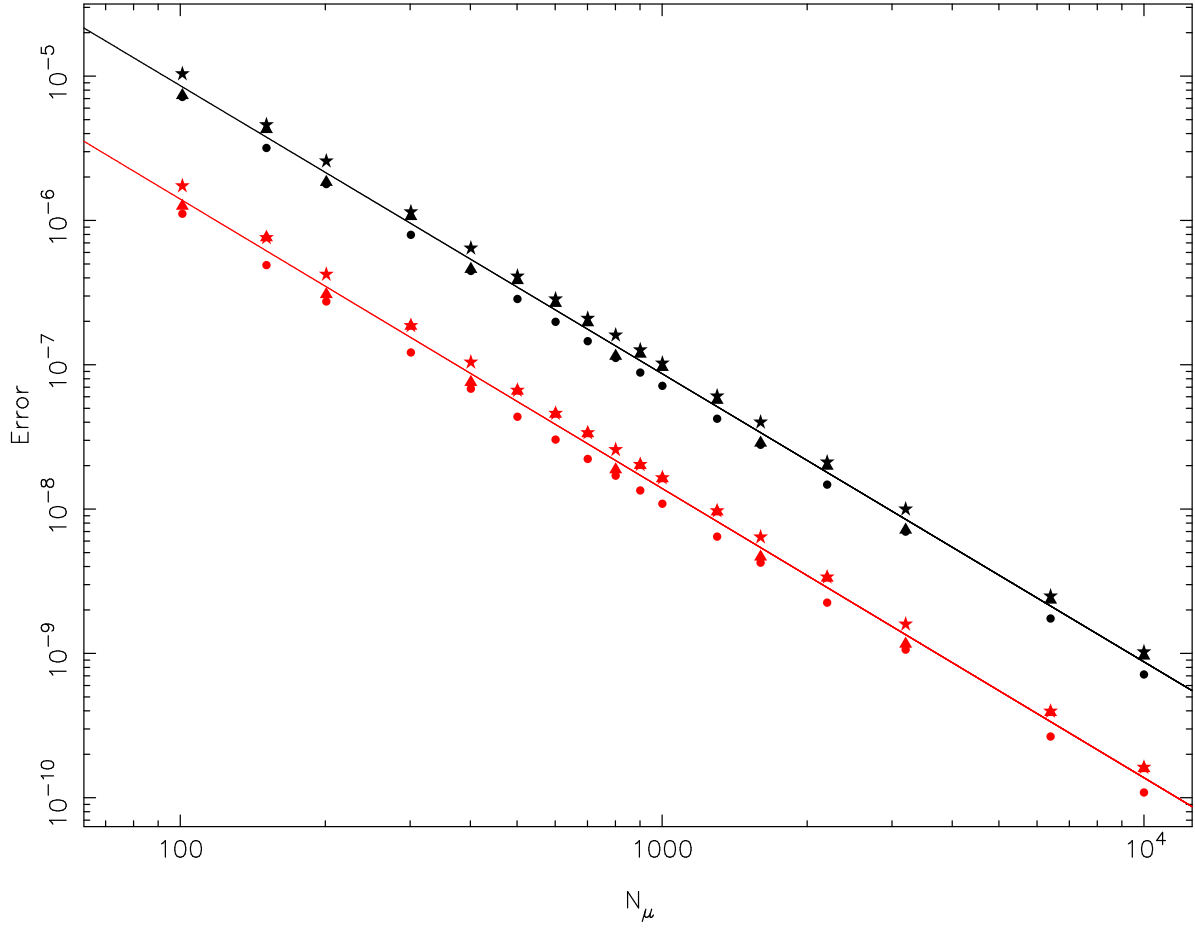}
\caption{The maximum error (black symbols) and the median
error (red symbols) in the flux fraction in the
$(p,z)$ plane as a function of the table
length $N_{\mu}$, where $p$ is the ratio of the radii and
$z$ is the distance between the centers in units of
the radius of the back body. The triangles are for the QL,
the circles are for the log law, and the stars are
for the square root law.}
\label{fig:plottableerror}
\end{figure}

\begin{figure}[t]
\includegraphics[angle=-90,scale=0.65]{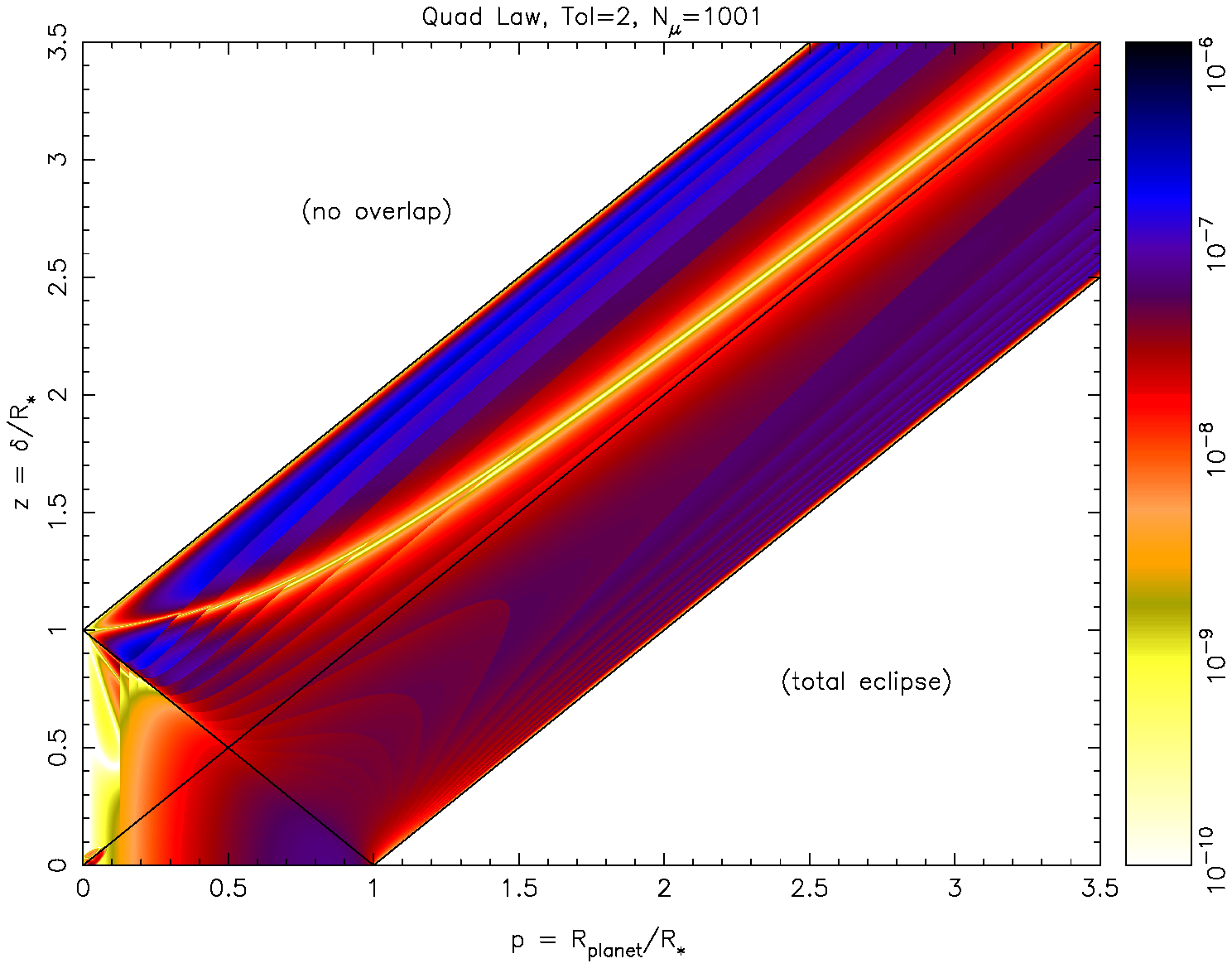}
\caption{The error in the flux fraction in the
$(p,z)$ plane, where $p$ is the ratio of the radii and
$z$ is the distance between the centers in units of
the radius of the back body.  This map was made using
$T_{\rm ATM}=2$ and a table length of
$N_{\mu}=1001$.}
\label{fig:error1}
\end{figure}

\begin{figure}[t]
\includegraphics[angle=-90,scale=0.65]{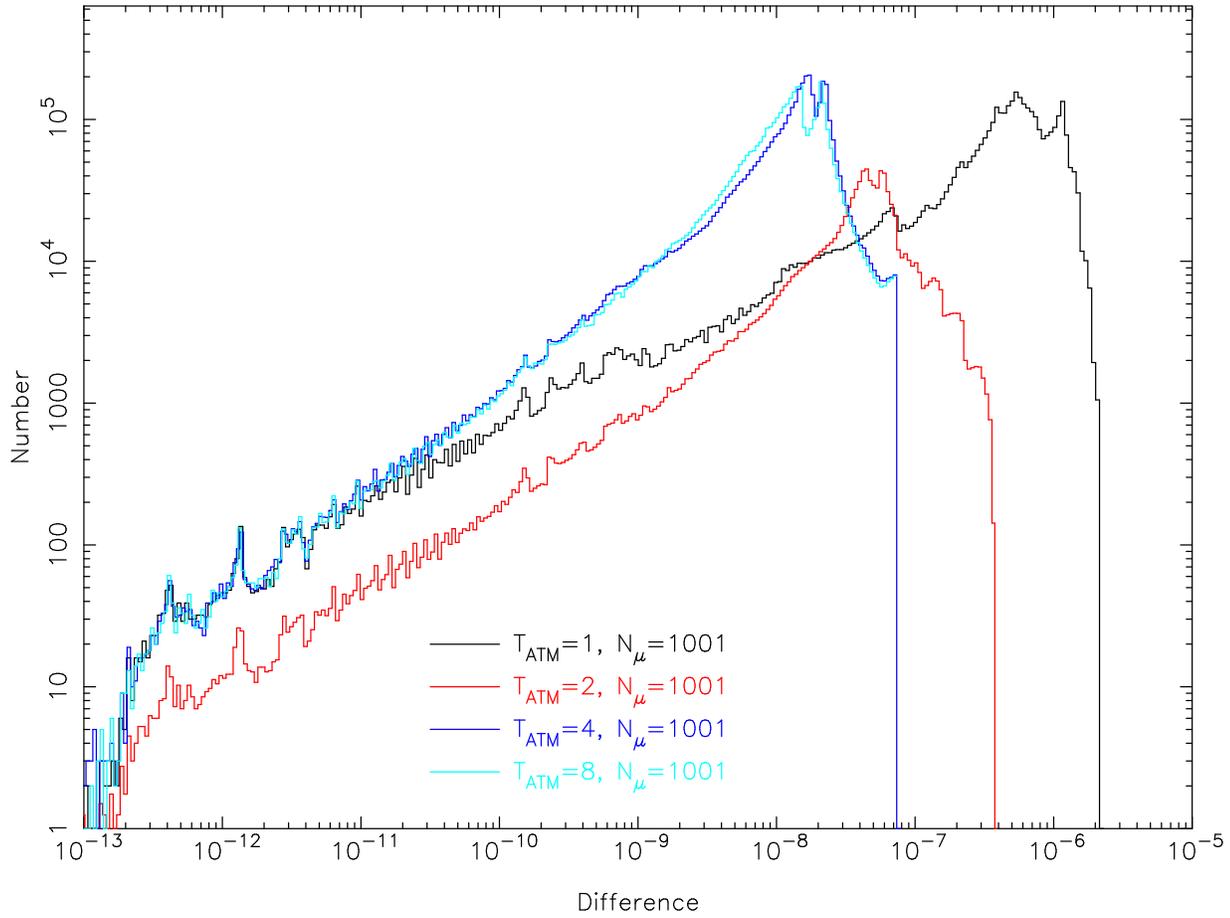}
\caption{The frequency distributions for the errors in the computation
of the flux fractions for various values of $T_{\rm ATM}$ with
a table length of $N_{\mu}=1001$.}
\label{fig:plothist1}
\end{figure}

\begin{figure}[t]
\includegraphics[angle=-90,scale=0.65]{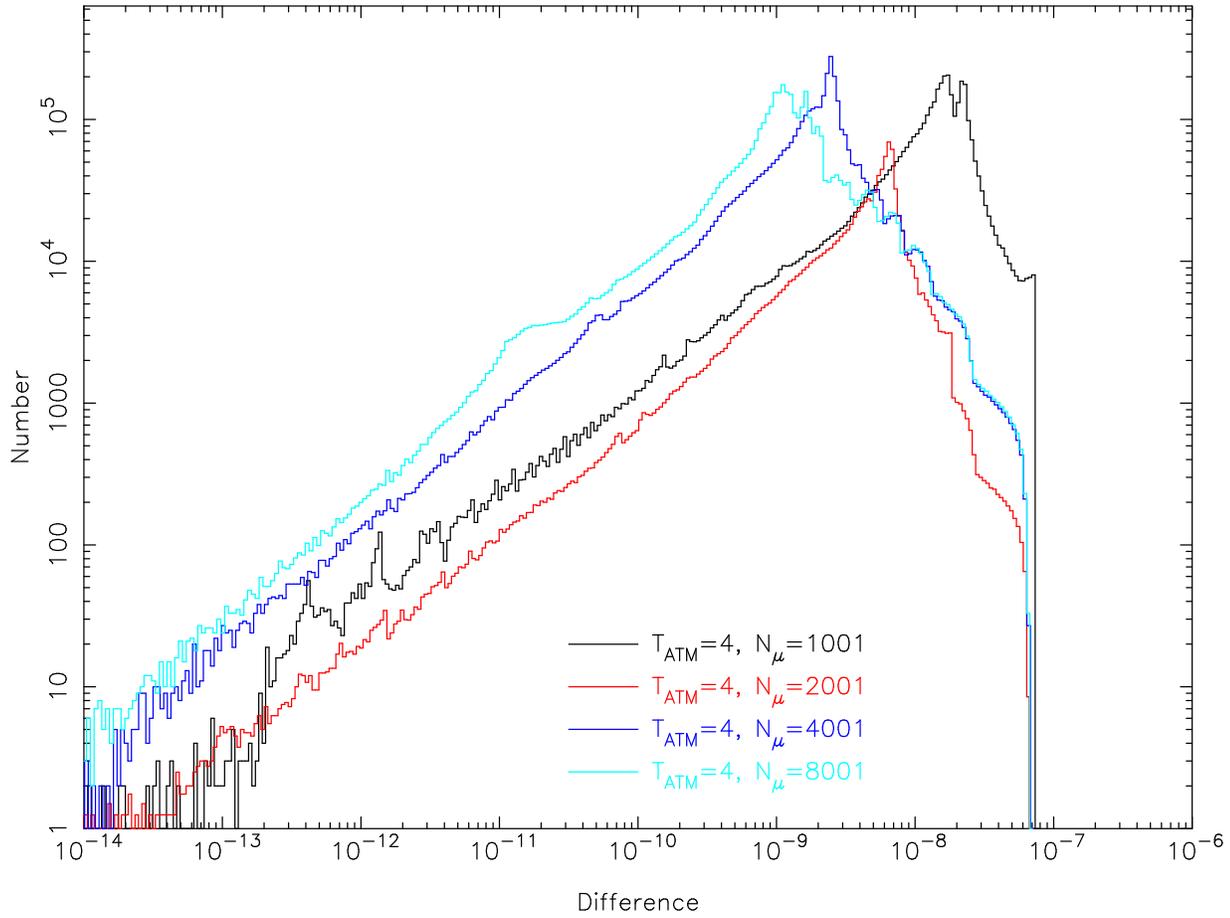}
\caption{The frequency distributions for the errors in the computation
of the flux fractions for various values of $N_{\mu}$ with
a tolerance of $T_{\rm ATM}=4$.}
\label{fig:plothist2}
\end{figure}

\begin{figure}[t]
\includegraphics[angle=-90,scale=0.65]{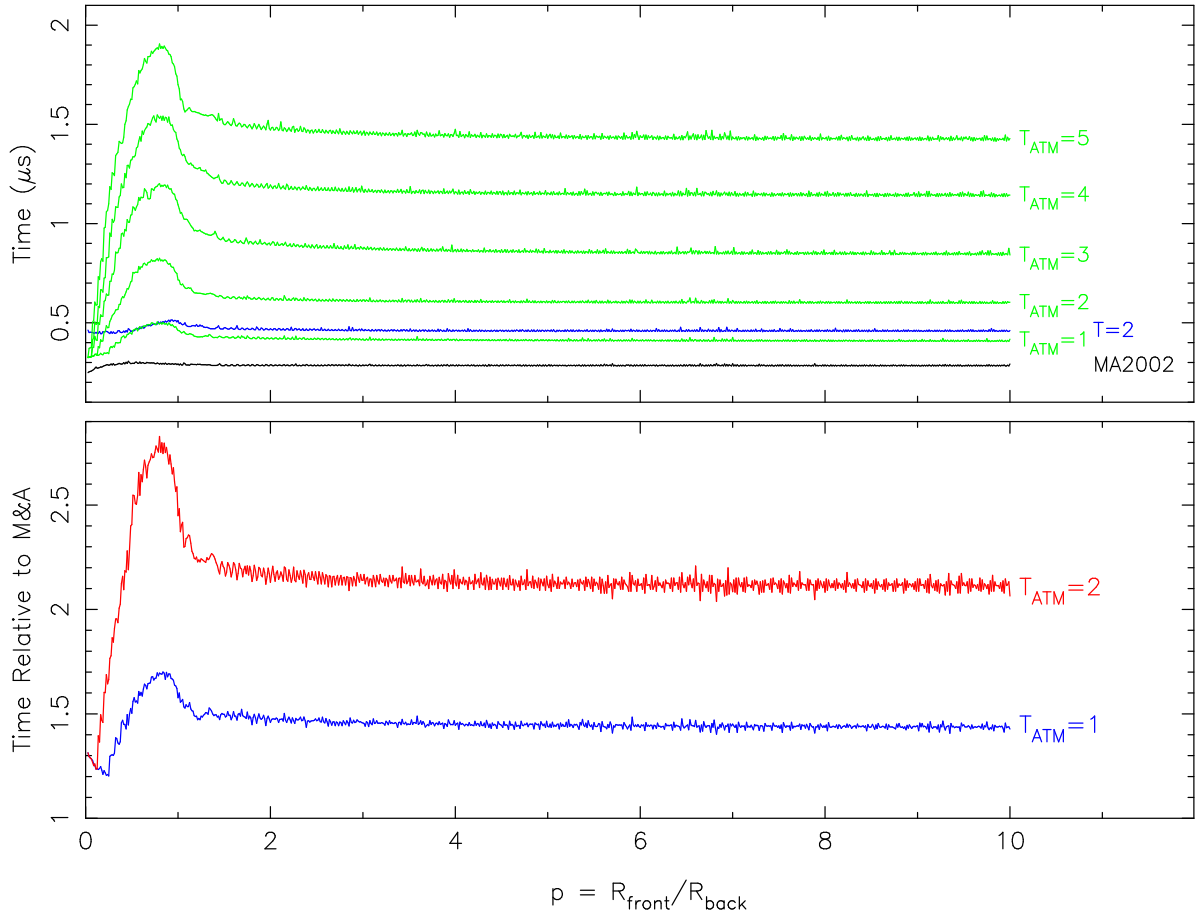}
\caption{Top: The time in $\mu$s needed to compute a flux
fraction as a function of the radius ratio $p$.
The black curve is for the {\tt occultquad} routine
from MA2002, and the blue curve is for the method
of ${\rm Paper~I}$, using T$=2$.  The green curves are
for the new method with $T_{\rm ATM}=1$, 2, 3, 4, and 5.
Bottom:  The ratio of times relative to the MA2002
{\tt occultquad} routine for $T_{\rm ATM}=1$ (blue)
and $T_{\rm ATM=2}$ (red).
}
\label{fig:plottiming}
\end{figure}

\begin{figure}[ht!]\
\plotone{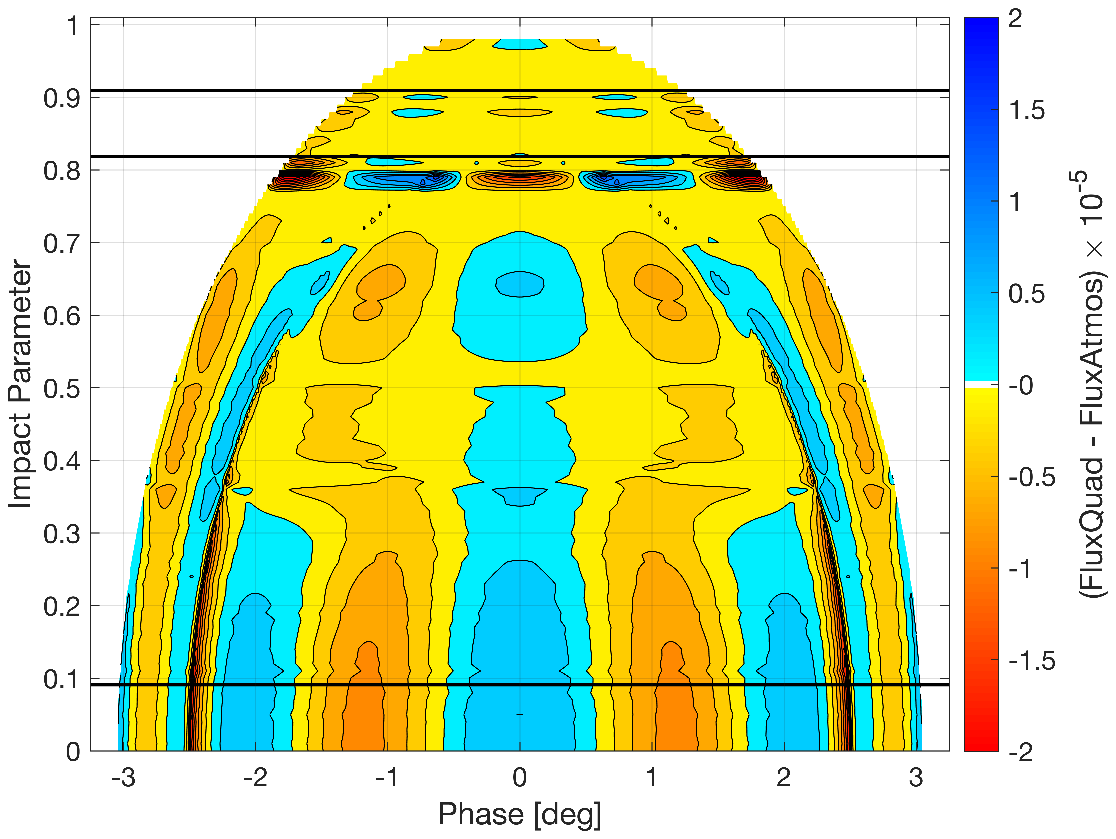}
\caption{As described in Section \ref{sec:hotjup}, there are 101 model transits light curves, 
each with the same stellar and planet radius constructed using the ATM.
The impact parameters of these models range from 0 to 1 in equal steps. Treating them as ``data'', each of these light curves was fitted using the code described in ${\rm Paper~I}$, with the analytic quad law. The fitting was done for 
four parameters, namely the two quad-law coefficients, the planet's radius, and the planet's orbital inclination. The 101 residual light curves are shown here stacked, where the color represents the size of the difference between the flux from the analytic fitting and that from the ATM. Blue represents a positive difference while red represents a negative difference. The lowest horizontal line represents the impact parameter for which the planet rim makes contact with 
the center of the star. The middle line represents the impact parameter for which the planet's rim makes contact with the inside 
of the stellar rim. The top line represents the parameter for which the planet's center makes contact with the stellar rim.}
\label{fig:FigAQuadKepDiff}
\end{figure}
\clearpage

\begin{figure}[ht!]\
\plotone{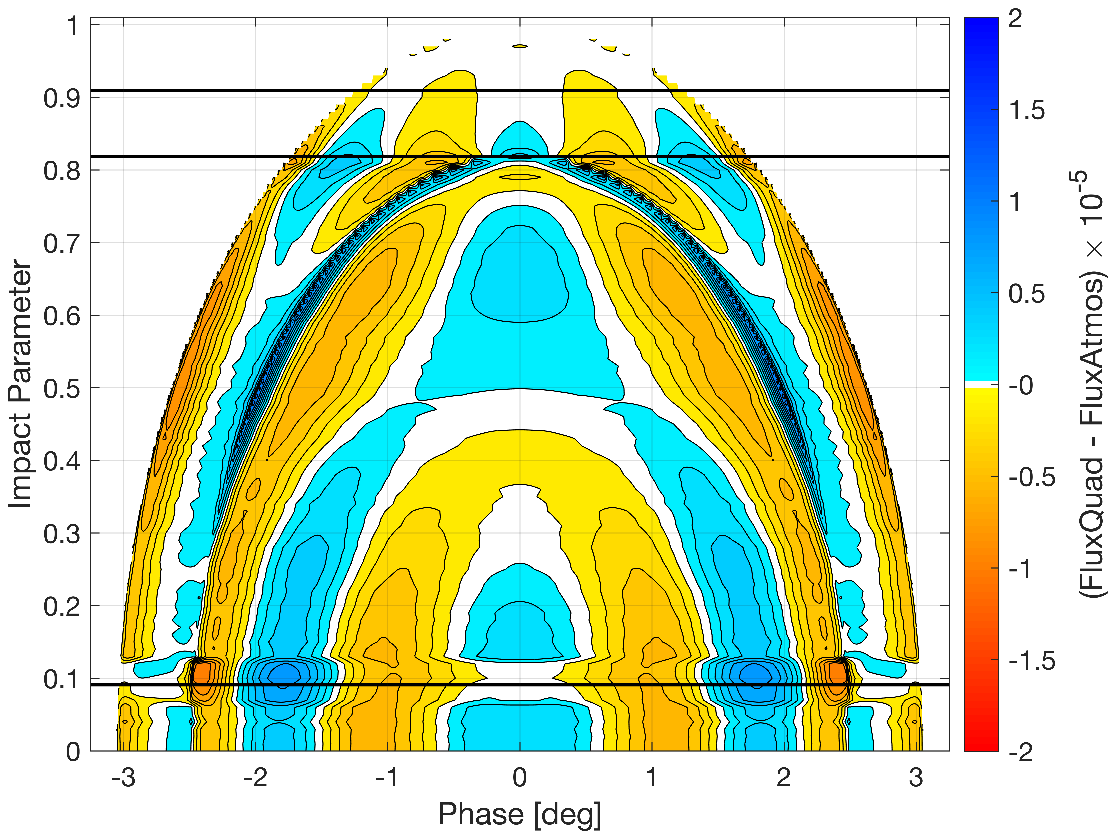}
\caption{The same construction as in Figure \ref{fig:FigAQuadKepDiff} but 
for the spectral B band.}
\label{fig:FigBQuadBDiff}
\end{figure}

\clearpage

\begin{figure}[ht!]\
\plotone{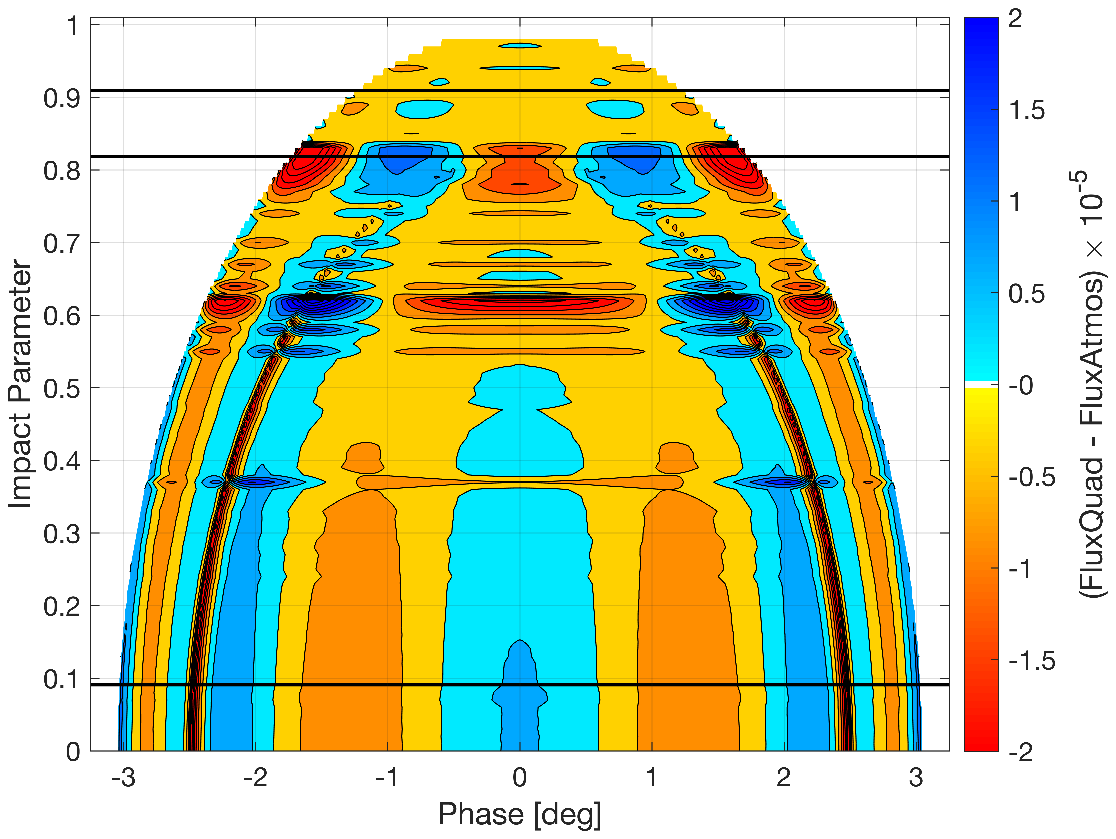}
\caption{The same construction as in Figure \ref{fig:FigAQuadKepDiff} but 
for the spectral H band.}
\label{fig:FigCQuadHDiff}
\end{figure}

\clearpage

\vspace*{5mm}
\begin{figure}[ht!]\
\hspace*{1cm}\includegraphics[angle=0,scale=0.8]{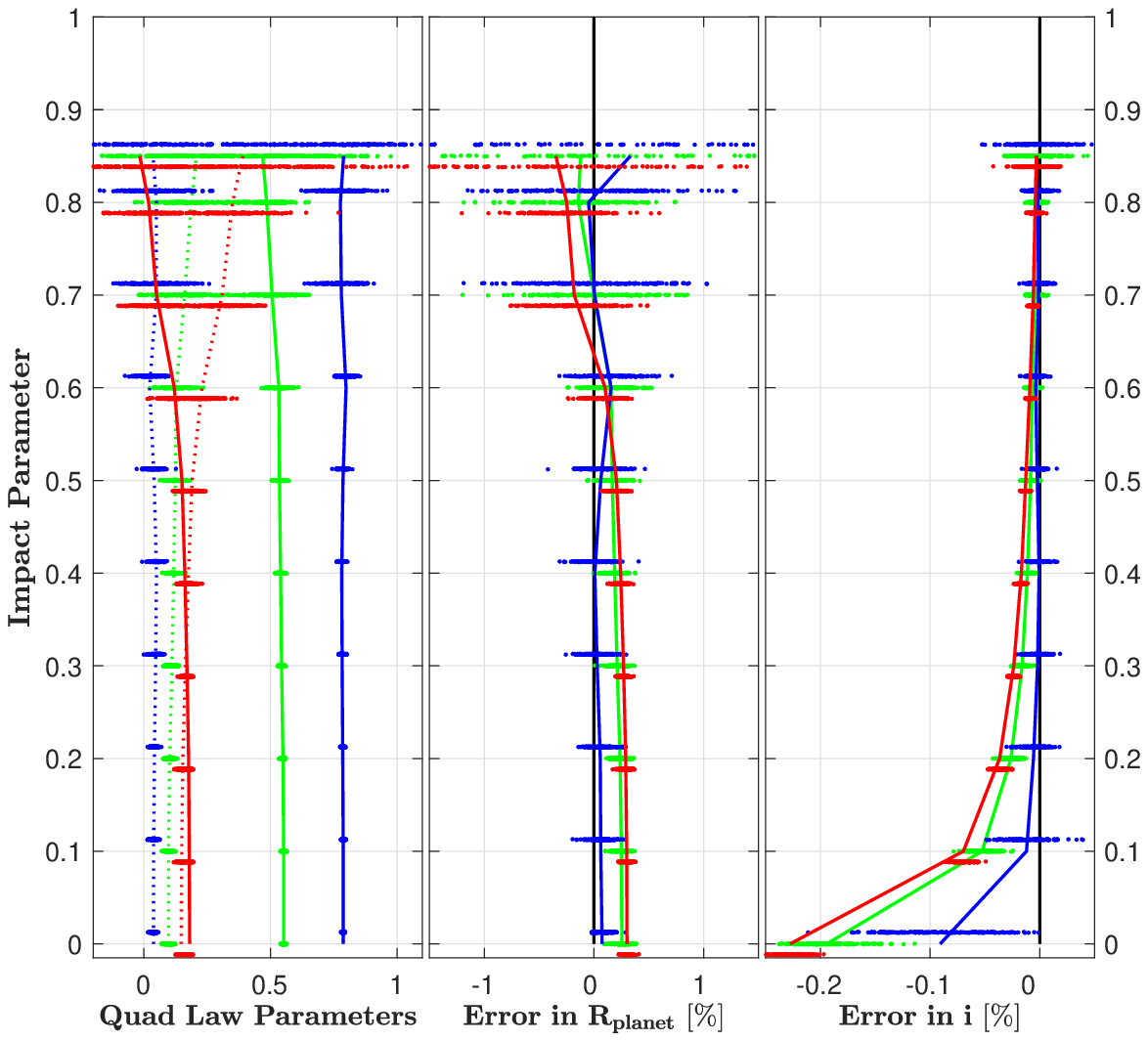}
\caption{The quad law coefficients $q_1$ (solid lines) and $q_2$ (dashed lines) 
derived from the fits in Figures \ref{fig:FigAQuadKepDiff}, \ref{fig:FigBQuadBDiff}, and 
\ref{fig:FigCQuadHDiff} are shown in the left panel as a function of the impact paramter where the colors represent the spectral bands, Kepler (green), B
(blue), and H (red). The middle panel shows the relative error in the radius, and the right hand panel shows the 
relative error in the inclination.
}
\label{fig:FigDQuad3Panel}
\end{figure}

\clearpage

\begin{figure}[ht!]\
\plotone{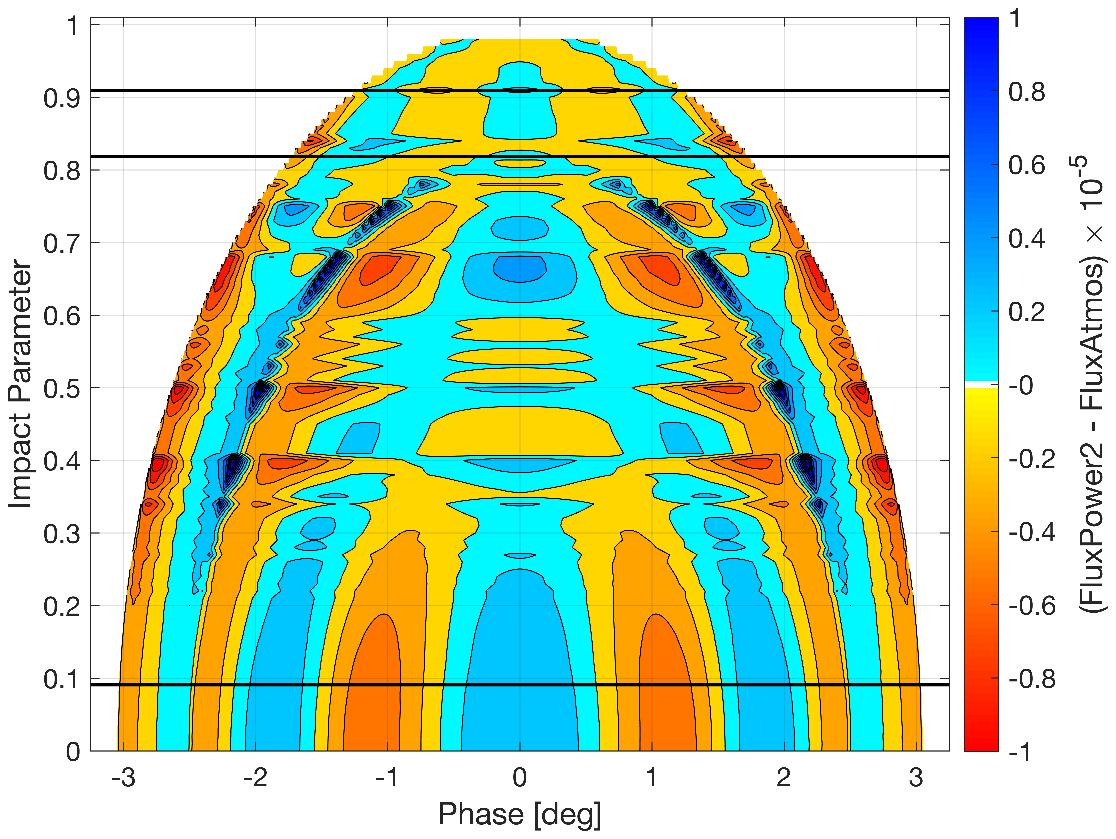}
\caption{The same construction as in Figure \ref{fig:FigAQuadKepDiff} but 
using the Power-2 law for the analytic model. The residuals for the Kepler spectral band
are shown.}
\label{fig:FigEPowKepDiff}
\end{figure}

\clearpage

\begin{figure}[ht!]\
\plotone{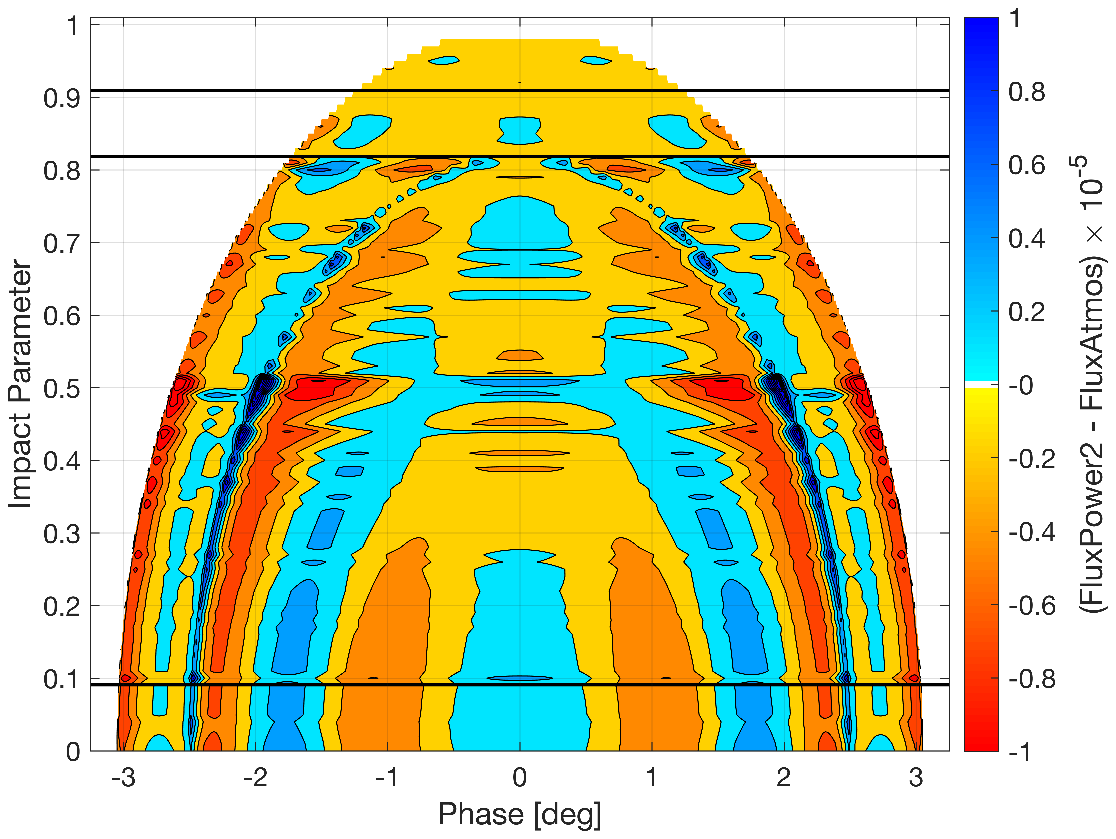}
\caption{The same construction, for the Power-2 law, as in Figure \ref{fig:FigEPowKepDiff} but  for the B spectral band.}
\label{fig:FigFPowBDiff}
\end{figure}

\clearpage

\begin{figure}[ht!]\
\plotone{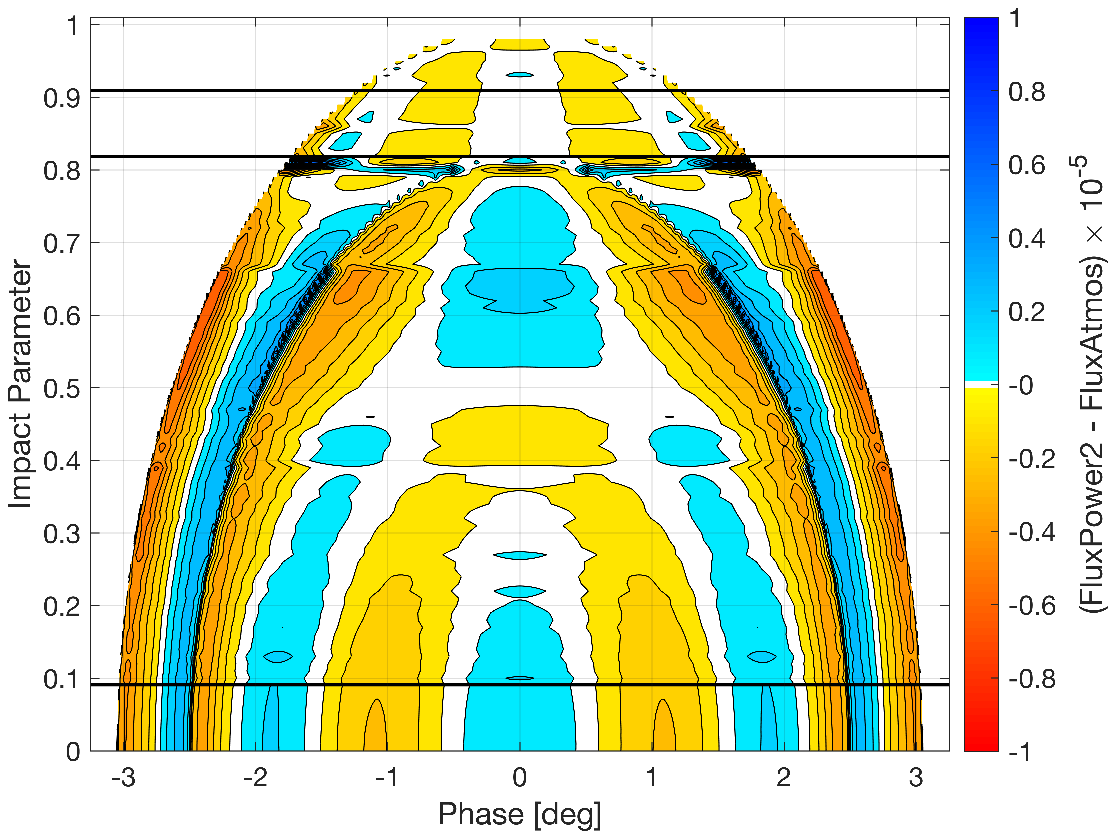}
\caption{The same construction, for the Power-2 law, as in Figure \ref{fig:FigEPowKepDiff} but for the H spectral band.}
\label{fig:FigGPowHDiff}
\end{figure}

\clearpage

\vspace*{5mm}
\begin{figure}[ht!]\
\hspace*{1cm}\includegraphics[angle=0,scale=0.8]{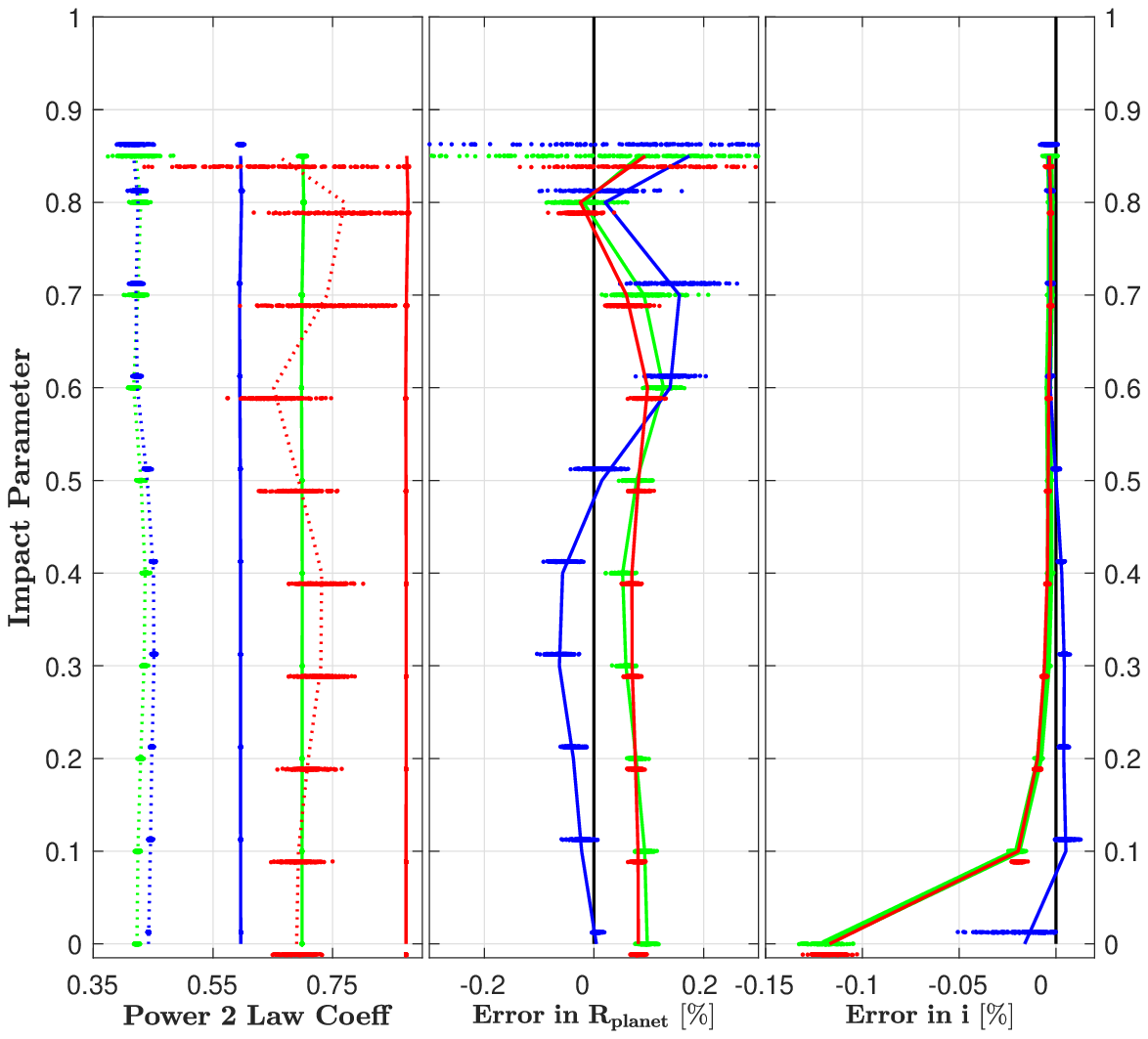}
\caption{Similar to Figure \ref{fig:FigDQuad3Panel}, but using the Power-2 law in the analytic model. Here, the left panel shows the optimal Maxted Power-2 Law coefficients, $h_1$ (solid lines), and $h_2$ (dashed lines). Since the first coefficient is defined as the value of the intensity at $\mu=0.5$, it has a very small variation with the impact parameter and with the optimization. (See also our discussion of these coefficients in \citealt{Short2019}.)
}
\label{fig:Pow2LawParamsAndErrors}
\end{figure}

\clearpage

\vspace*{5mm}
\begin{figure}[ht!]\
\hspace*{1cm}\includegraphics[angle=0,scale=0.8]{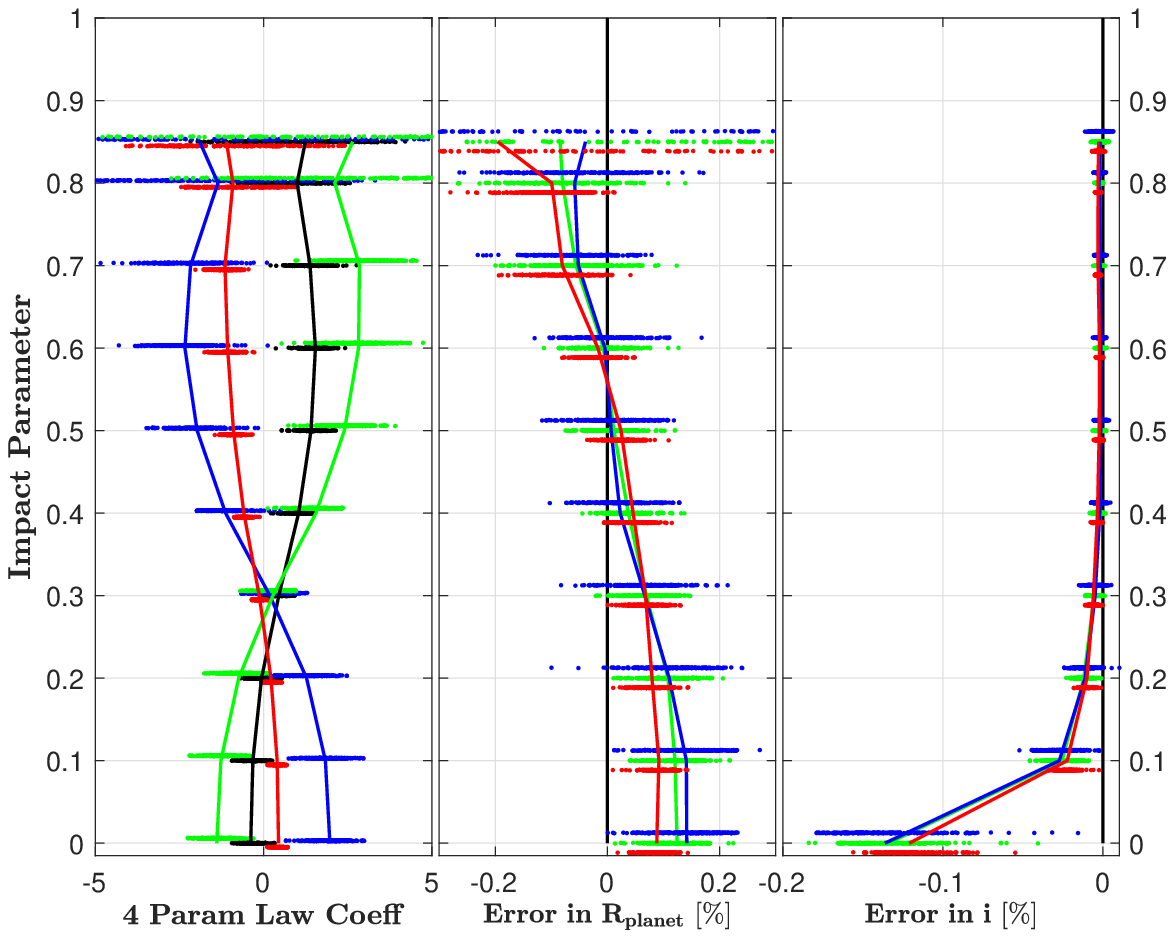}
\caption{Similar to Figure \ref{fig:FigDQuad3Panel}, but using the 4-parameter (``Claret'') law in the analytic model. 
The colors in the middle and right panels represent the spectral bands as before. The left panel shows the optimized coefficients for the Kepler band only: $c_1$ (black), $c_2$ (blue), $c_3$ (green), and $c_4$ (red).
}
\label{fig:Claret3PanelNov28}
\end{figure}

\clearpage

\begin{figure}[ht!]\
\plotone{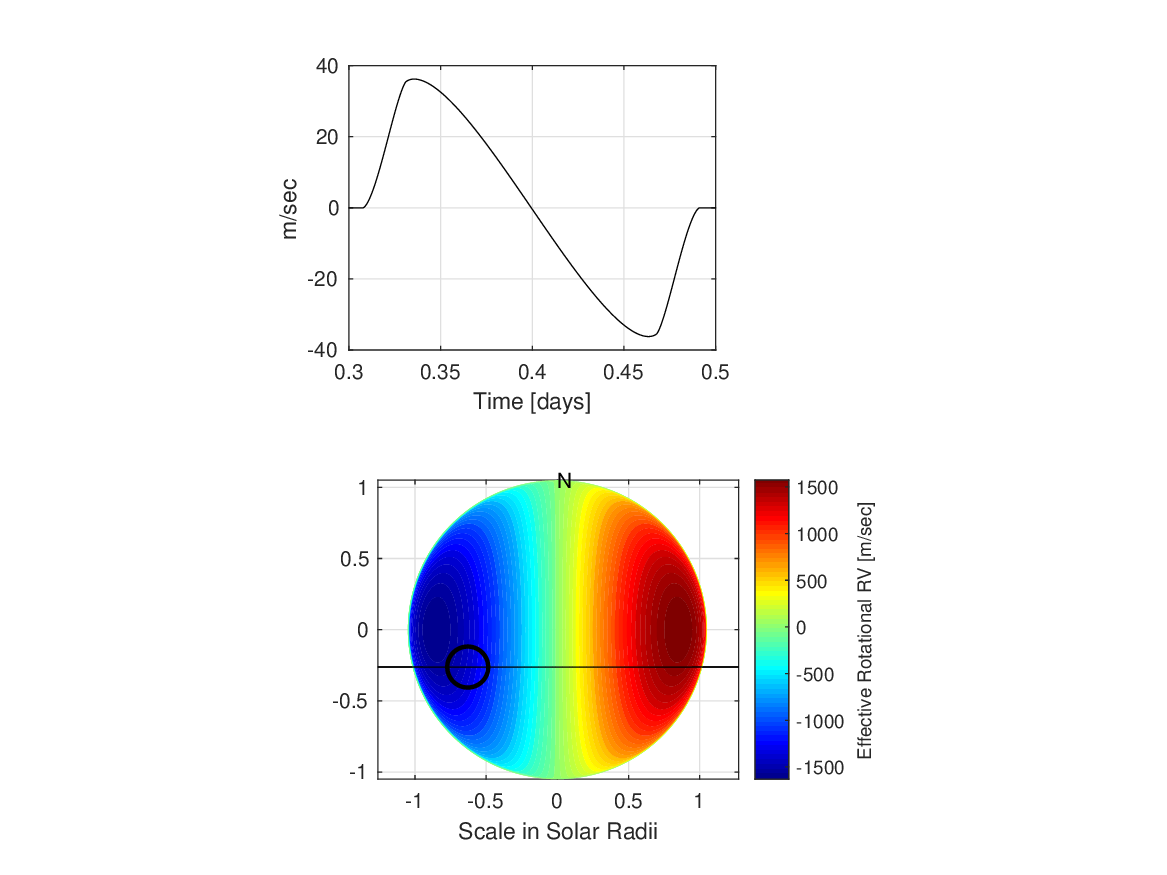}
\caption{Top: the change in the effective rotational radial velocity
(R-M Effect) for the Hot Jupiter example with an impact parameter of
0.25 and using an inclination of $89.2545^\circ$, Rotation Period of
20 days, $\Phi_{\rm rot} = 90^\circ$ , and a $\Theta_{\rm
  rot}=0^\circ$.  Bottom: color represents the effective rotational
radial velocity field on the star scaled by the color bar.  The
black circle is the limb of the transiting planet and the black line
is the orbital track of the planet across the POS view of the
star. $N$ is the north pole of the right handed stellar rotation
axis.
}
\label{fig:RMEffect90_0HotJupIP25}
\end{figure}

\begin{figure}[ht!]\
\plotone{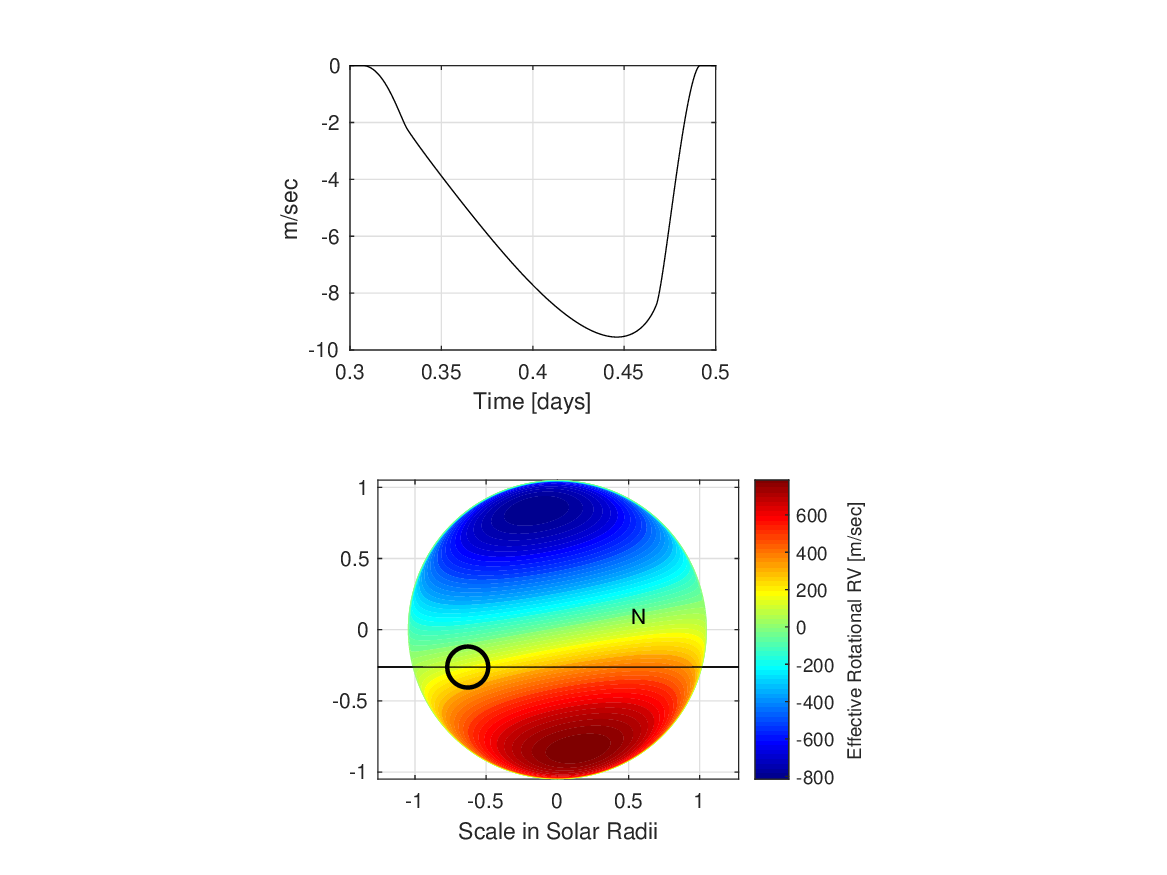}
\caption{Similar to Figure \ref{fig:RMEffect90_0HotJupIP25} but with a 
misaligned rotation axis given by $\Phi_{\rm rot} = 30^\circ$, and 
$\Theta_{\rm rot}=-80^\circ$.}
\label{fig:RMEffect30_m80HotJupIP25}
\end{figure}

\begin{figure}[ht!]\
\plotone{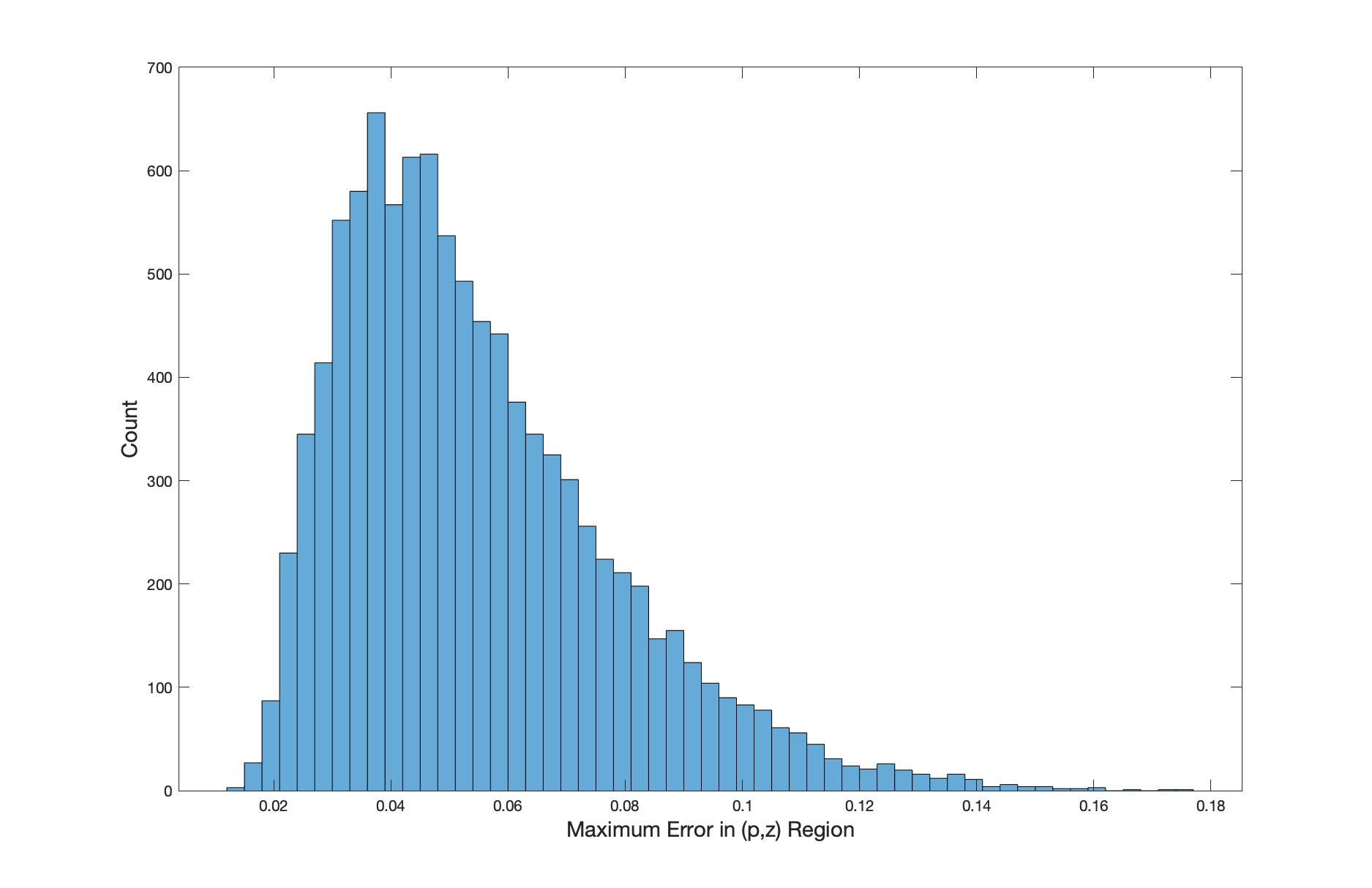}
\caption{Histogram of the $\max_{\rm p~and~q}|F_U(p,z)|$ computed
for 10,000 sample error tables using the unit
normal distribution over the range in the $(p,z)$
plane given by $0<p<3$ and $0<z<3$.  }
\label{fig:DistriMaxErrorUnitNormal}
\end{figure}

\begin{figure}[ht!]\
\plotone{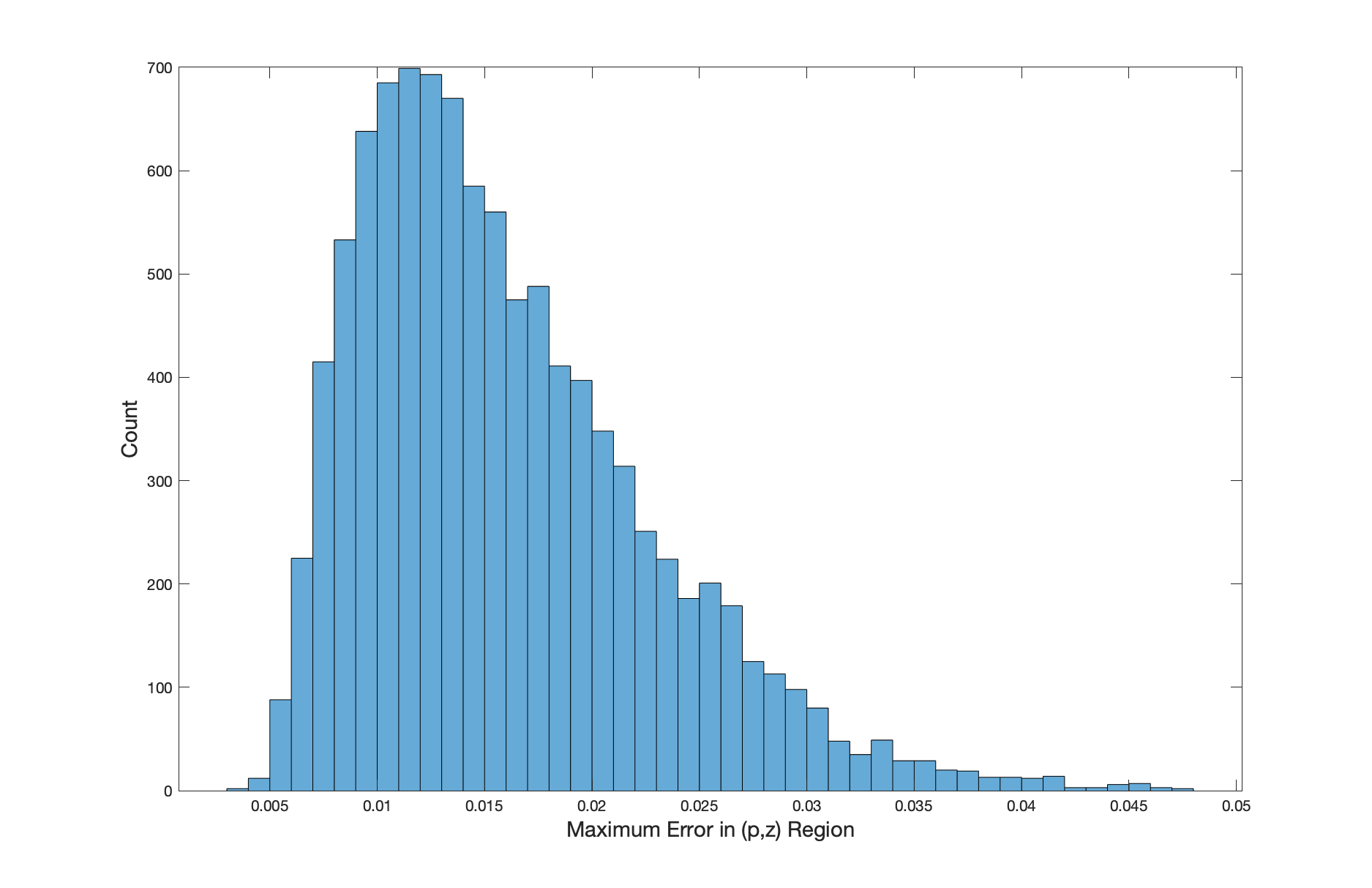}
\caption{Histogram of the $\max_{\rm p and q}|F_U(p,z)|$ computed
for 10,000 sample error tables using the unit
uniform distribution over the range in the $(p,z)$
plane given by $0<p<3$ and $0<z<3$. }
\label{fig:DistriMaxErrorUnitUniform}
\end{figure}

\begin{figure}[ht!]\
\plotone{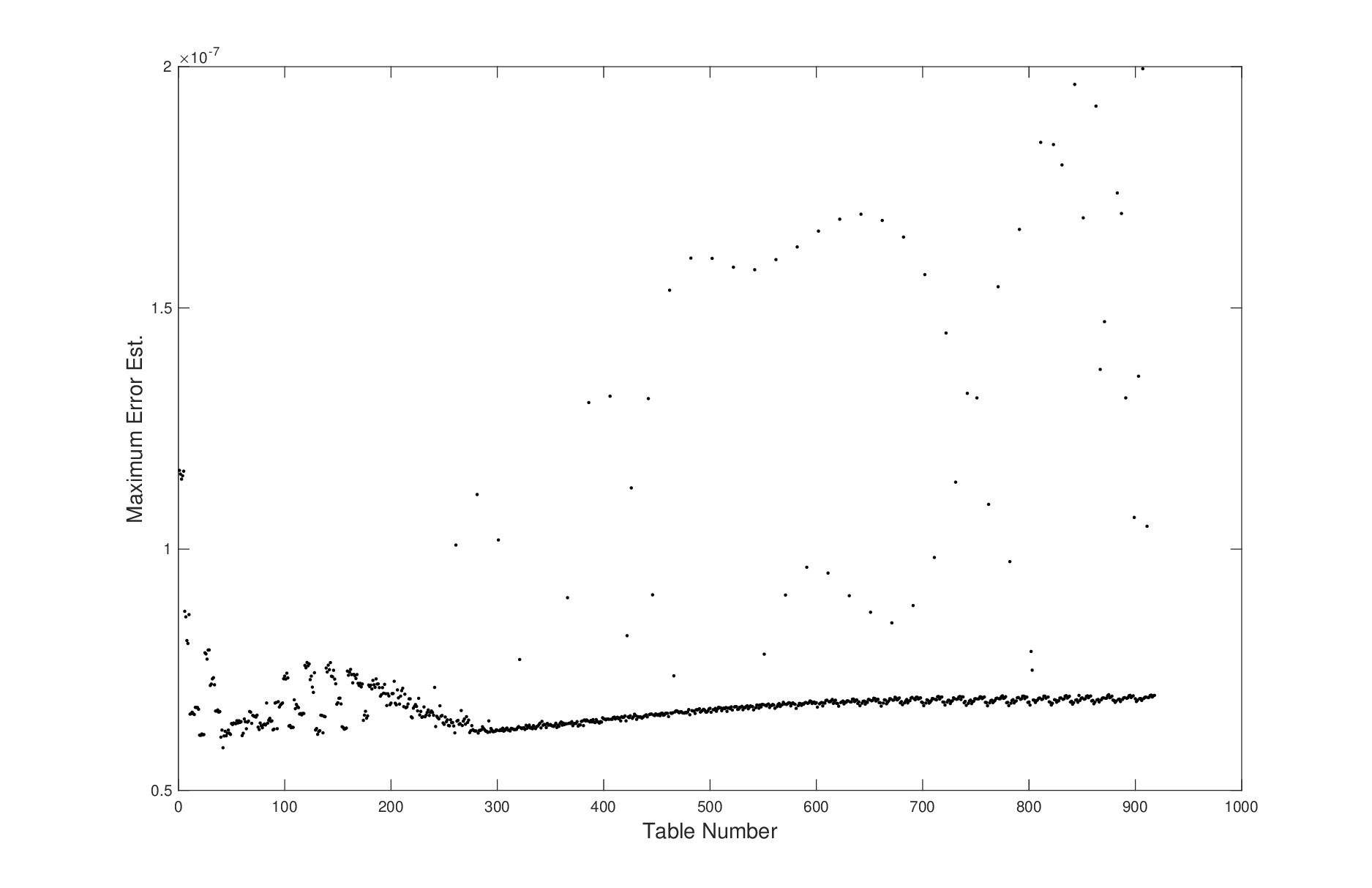}
\caption{The maximum difference between the PL
and PCH function approximations computed over
 in the region $0<p<3$
and $0<z<3$ in the $(p,z)$ plane for each of
the 918 models in the \citet{Neilson2013} grid is shown.
The model file {\tt t7500g475m5} was omitted from
the analysis since it was an outlier.}
\label{fig:MaxLCErrorforNandLTables}
\end{figure}

\begin{figure}[ht!]\
\plotone{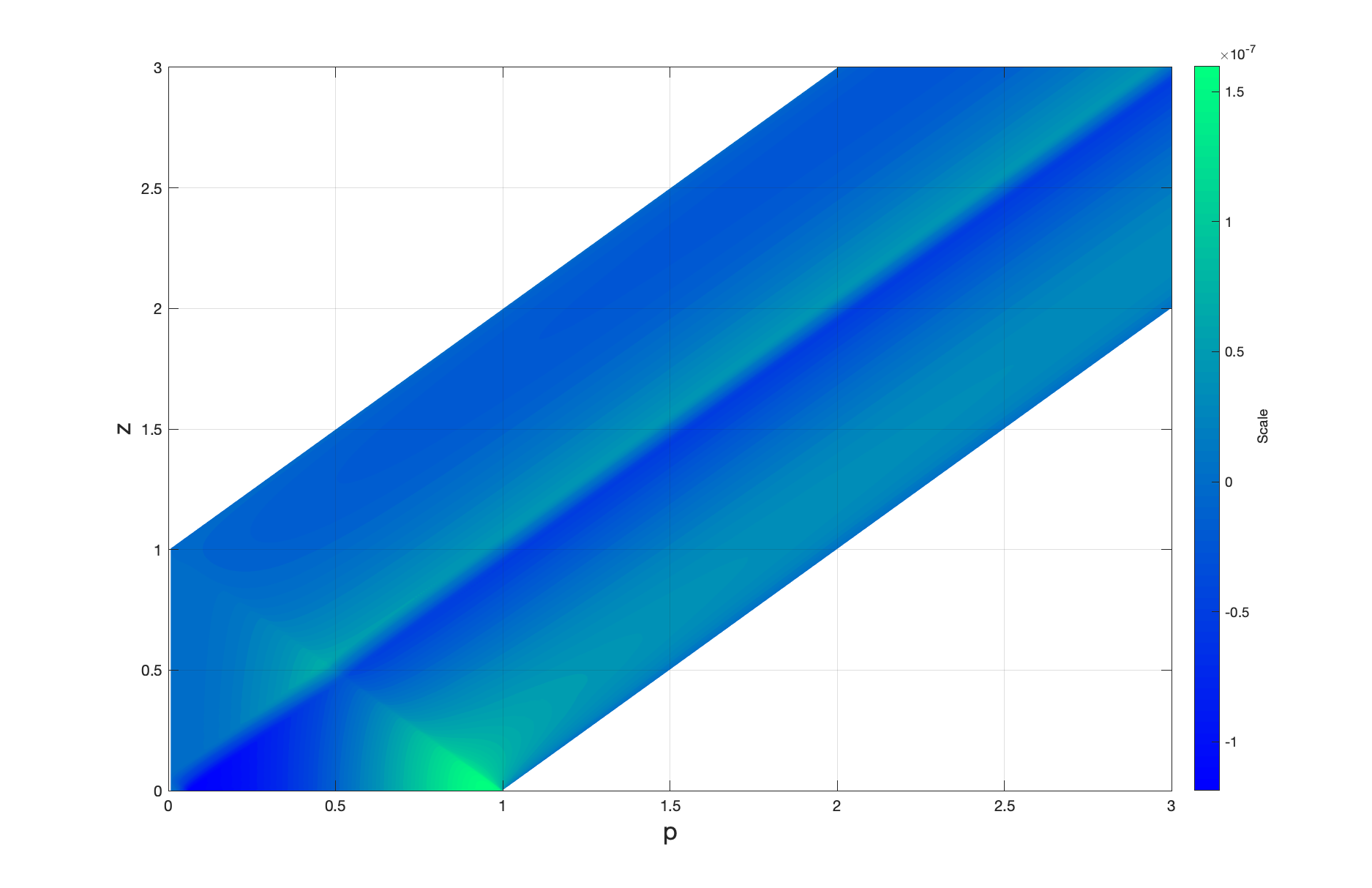}
\caption{Error estimate in the $(p,z)$ plane for the 
\citet{Neilson2013} table for $T_{\rm
  eff}=5700,~\log(g)=4.50$, and mass $M=1.1M_\odot$. }
\label{fig:Error10kminus1000t5700g450m11}
\end{figure}

\end{document}